\documentclass[usenatbib]{mnras} 
\usepackage{newtxtext,newtxmath}
\usepackage[T1]{fontenc}
\usepackage{ae,aecompl}
\usepackage{ulem}
\usepackage{color}
\usepackage{graphicx}
\usepackage{amsmath}
\usepackage{mathrsfs}
\usepackage{xfrac}
\usepackage{multirow}
\usepackage[flushleft]{threeparttable}
\newif\ifAMStwofonts
\AMStwofontstrue
\newcommand{\Spitzer}{{\it Spitzer }}
\newcommand{\lsun}{L$_{\odot}$}
\newcommand{\msun}{M$_{\odot}$}
\newcommand{\msunrate}{M$_{\odot}$\,yr$^{-1}$}
\newcommand{\mstar}{$M^*$}
\newcommand{\CII}{\mbox{{\sc [C~ii]}}}
\newcommand{\MgII}{\mbox{[Mg~{\sc ii}]}}
\newcommand{\OII}{\mbox{[O~{\sc ii}]}}
\newcommand{\OIII}{\mbox{O~{\sc iii}}}

\title[SPT2349-56 at optical wavelengths]{Optical and near-infrared observations of the SPT2349-56  proto-cluster core at \textbf{\textit{z}} = 4.3
} 
\author[K.\ M.\ Rotermund et al.]
{K.\ M.\ Rotermund$^{1}$\thanks{kaja@dal.ca}, 
S.\ C.\ Chapman$^{1,2,3}$, 
K.\ A.\ Phadke$^4$,
R.\ Hill$^3$, E.\ Pass$^{5}$, M.\ Aravena$^{6}$,
\newauthor M.\ L.\ N.\ Ashby$^5$, A.\ Babul$^7$, M.\ B\'{e}thermin$^8$, R.\ Canning$^9$, C.\ de Breuck$^{10}$, C.\ Dong$^{11}$, 
\newauthor A.\ H.\ Gonzalez$^{11}$, 
C.\ C.\ Hayward$^{12}$, S.\ Jarugula$^4$, D.\ P.\ Marrone$^{13}$, D.\ Narayanan$^{11}$, 
\newauthor C.\ Reuter$^4$, D.\ Scott$^3$,
J.\ S.\ Spilker$^{14,\dag}$,
J.\ D.\ Vieira$^4$, 
G.\ Wang$^3$, A.\ Weiss$^{15}$
\\
$^{1}$Department of Physics and Atmospheric Science, Dalhousie University, Halifax, NS, B3H 4R2, Canada\\
$^{2}$National Research Council, Herzberg Astronomy and Astrophysics, 5071
West Saanich Road, Victoria, V9E 2E7, Canada\\
$^{3}$Department of Physics and Astronomy, University of British Columbia, Vancouver, BC, V6T 1Z1, Canada\\
$^4$Department of Astronomy, University of Illinois, 1002 West Green Street, Urbana, IL 61801, USA\\
$^{5}$Harvard-Smithsonian Center for Astrophysics, 60 Garden Street, Cambridge, MA 02138, USA\\
$^{6}$N\'{u}cleo de Astronom\'{\i}a, Facultad de Ingenier\'{\i}a y Ciencias, Universidad Diego Portales, Santiago, Chile.\\
$^7$Department of Physics and Astronomy, University of Victoria, Victoria, BC, V8P 5C2, Canada\\
$^8$Aix-Marseille Universit\'{e}, CNRS, LAM, Laboratoire d’Astrophysique de Marseille, Marseille, France\\
$^9$Stanford University,
382 Via Pueblo Mall 
Stanford, CA 94305-4013\\
$^{10}$European Southern Observatory, Karl Schwarzschild Stra\ss e 2, 85748 Garching, Germany \\
$^{11}$Department of Astronomy, University of Florida, Gainesville, FL 32611, USA\\
$^{12}$Center for Computational Astrophysics, Flatiron Institute, 162 Fifth Avenue, New York, NY 10010, USA\\
$^{13}$Steward Observatory, University of Arizona, 933 North Cherry Avenue, Tucson, AZ 85721, USA\\
$^{14}$Department of Astronomy, University of Texas at Austin, 2515, USA\\
$^{15}$Max-Planck-Institut für Radioastronomie, Auf dem Hügel 69 D-53121 Bonn, Germany\\
$^{\dag}$NHFP Hubble Fellow
}

\date{Accepted XXX. Received YYY; in original form ZZZ}

\pubyear{2020}

\begin{document}
\label{firstpage}
\pagerange{\pageref{firstpage}--\pageref{lastpage}}
\maketitle

\begin{abstract} 
We present Gemini-S and {\it Spitzer}-IRAC optical-through-near-IR observations in the field of the SPT2349-56 proto-cluster at $z=4.3$. We detect optical/IR counterparts for only nine of the 14 submillimetre galaxies (SMGs) previously identified by ALMA in the core of SPT2349-56. In addition, we detect four $z\sim4$ Lyman-break galaxies (LBGs) in the 30\,arcsec-diameter region surrounding this proto-cluster core. Three of the four LBGs are new systems, while one appears to be a counterpart of one of the nine observed SMGs. We identify a candidate brightest cluster galaxy (BCG) with a stellar mass of \mbox{$(3.2^{+2.3}_{-1.4})\times10^{11}$\,\msun}. 
The stellar masses of the eight other SMGs place them on, above, and below the main sequence of star formation at $z\approx4.5$. The cumulative stellar mass for the SPT2349-56 core is at least \mbox{$(12.2\pm2.8)\times10^{11}$\,\msun}, a sizeable fraction of the stellar mass in local BCGs, and close to the universal baryon fraction (0.19) relative to the virial mass of the core ($10^{13}$\,\msun). As all 14 of these SMGs are destined to quickly merge, we conclude that the proto-cluster core has already developed a significant stellar mass at this early stage, comparable to $z=1$ BCGs. Importantly, we also find that the SPT2349-56 core structure would be difficult to uncover in optical surveys, with none of the ALMA sources being easily identifiable or constrained through $g,r,$ and $i$ colour-selection in  deep optical surveys and only a modest overdensity of LBGs over the more extended structure. SPT2349-56 therefore represents a truly dust-obscured phase of a massive cluster core under formation.
\end{abstract}


\section{Introduction}

Submillimetre galaxies (SMGs), which are forming stars at prodigious rates, even sometimes exceeding $1{,}000$\,\msunrate\ \citep[e.g.][]{Vieira13,Swinbank13}, 
are prominent during the peak of galaxy assembly at $z>2$ \citep{Chapman03N, Chapman05} and likely play an important role in the history of early massive galaxy formation, with a high-$z$ tail still dominant at $z\simeq4{-}5$ 
(e.g. \citealt{Weiss}; \citealt{Reuter20}).
SMGs at $z\simeq2.5$ often have stellar masses on the order of $10^{11}$\,\msun\ \citep{Hainline11,Ma15}. Many are 
found significantly above the `main sequence' (MS) of star-forming
galaxies (in specific star formation rate; e.g. \citealt{Hainline11}), but are also convincingly found on the massive end of the main sequence (e.g., \citealt{Michalowski12}). 
Their rapid evolution early in cosmic time stresses the importance of understanding SMGs at high redshift, with current models still struggling to match their detailed properties \citep[e.g.][]{Cowley}.

SMGs have recently been directly identified as an important star-formation mode in overdense proto-clusters of galaxies in the early Universe \citep[e.g.][]{Chapman09,Casey16,Miller,Oteo18,Umehata19,Lacaille19}.
SMGs can be sites of intense star formation often long before the height of galaxy assembly \citep[e.g.][]{Casey} when a much larger fraction of star formation was occurring in overdense, collapsing proto-clusters of galaxies \citep{Chiang17}.  Observing massive SMGs at the highest redshifts is therefore crucial for understanding the evolution of large-scale structures.  Additionally, galaxy proto-clusters are interesting laboratories in which the mass budget of galaxies in dense environments can be studied. 
Identifying differences between field SMGs and those growing within overdensities, such as in proto-cluster cores, can help elucidate aspects of galaxy evolution that lead to the vastly different properties of galaxies found 
in clusters and in the field at the present epoch. 
SMGs growing in the dense environments of proto-clusters are expected to have formed earlier, be more massive, and undergo major-mergers more frequently than their field galaxy counterparts \citep{Overzier,Rennehan}. 
Furthermore, the enormous early build-up of mass in galaxy proto-clusters makes them critical when investigating large-scale structures in the Universe and can potentially help constrain cosmological parameters \citep[e.g.][]{Wen}. 

Early high-redshift proto-cluster discoveries arose from spectroscopic studies of Lyman-break galaxies (LBGs), 
notably the spectroscopic confirmation of the $z=3.09$ proto-cluster in the SSA22 field by \cite{Steidel98}. 
This was followed by the detection of the $z=2.3$ proto-cluster in the field of the QSO HS1700+643 \citep{Steidel05}, using a modification of the LBG technique for $z\,{\approx}\,2$ galaxies.
The \cite{Steidel98} colour selection technique, which identifies LBGs within a specific redshift range, has been adopted and modified successfully by many subsequent studies, and has been used to search for overdensities over wide fields. Prominent recent examples are the identification of 179 proto-cluster candidates at $z\approx3.8$ from the Hyper Suprime-Cam Subaru  program by \cite{Toshikawa18}. Strategic searches for overdensities of Ly$\alpha$ emitters or H$\alpha$ emitters near objects strongly suspected to be the progenitors of massive galaxies at cluster cores, i.e. high-$z$ radio galaxies and QSOs, have also proved successful in identifying proto-clusters. Some notable examples are the discovery of excess Ly$\alpha$ emitters, H$\alpha$ emitters, and 
extremely red objects in the field of the radio galaxy MRC1138–262 at $z=2.16$ \citep{Pentericci00,Kurk04}, and the large-scale structure of Ly$\alpha$ emitters at $z=4.86$ \citep{Shimasaku03}, at $z=6.01$  \citep{Toshikawa12}, and at $z=5.7\,\&\,6.6$ \citep{Higuchi19,Harikane19}, all 
in the Subaru Deep Field. Some have also been discovered serendipitously with confirmed spectroscopic redshifts obtained in the VIMOS Ultra-Deep Survey \citep{Fevre15}, such as the $z=2.45$ \citep{Cucciati18} and $z=4.57$ \citep{Lemaux18} proto-clusters, both in the COSMOS field. 

More recently, so-called extended Ly$\alpha$ `blobs' and SMGs have been suggested as good tracers of overdense regions at high redshift \citep[\& references therein]{Overzier}. 
In \cite{Vieira10}, it was proposed that a population of bright millimetre-selected sources identified in wide field surveys conducted by the South Pole Telescope (SPT) might contain a subset of unlensed, extremely luminous galaxies or groups of galaxies. This motivated a search for proto-cluster candidates in the SPT survey.

SPT uncovered a population of dusty, thermal sources selected at millimetre wavelengths \citep{Vieira10, Mocanu, Everett20}, which were predominantly identified through Atacama Large Millimeter/submillimeter Array (ALMA) imaging and spectroscopy to be gravitationally lensed SMGs at $z>3$ \citep{Vieira13}. However, detailed lens modelling of the population revealed several examples that appeared to {\it not} have significant lensing magnification \citep{Spilker} as well as lack  bright foreground lensing galaxies even in deep imaging (Rotermund in prep.). These {\it unlensed} sources are candidate proto-cluster cores (Wang et al. submitted, Chapman in prep.), of which the now well-studied $z=4.3$ SPT2349-56 \citep{Miller, Hill20} represents the brightest example in this SPT proto-cluster (SPT-PC) survey. 
High-redshift proto-cluster candidates have also been identified in {\it Herschel} surveys \citep{Lewis18}; one $z=4$ candidate has been followed up with ALMA and confirmed as a masssive proto-cluster core by \cite{Oteo18}, with a follow-up study of the member galaxies presented in \cite{Long20} and \cite{Ivison20}.

SPT2349-56 was detected as a thermal dust source, slightly resolved even by the 1-arcmin SPT beam. 
At 1.4\,mm, one of the three SPT bands, it has a peak flux density of $S_{1.4\,\mathrm{mm}}=23.3$\,mJy \citep{Miller},
comparable to the mean deboosted flux density of the SPT-SMG sample of 23\,mJy \citep{Reuter20}. 
Follow-up observations were initially conducted at 870\,$\mu$m with the Large APEX BOlometer CAmera (LABOCA, \citealt{APEX}) on the Atacama Pathfinder Experiment (APEX) telescope in order to obtain a more precise location on the sky. At LABOCA's 19-arcsec resolution, SPT2349-56 was resolved into two elongated sources, in contrast to the majority of the SPT sample, which continue to appear as unresolved sources at this resolution, singling it out as a possible extended structure of galaxies. The bright southern source was found to have a flux density of $S_{870\,\mu\mathrm{m}}\approx77$\,mJy \citep{Miller} and is clearly the locus of activity and centre-of-mass of the proto-cluster system \citep{Hill20}. Surrounding structures include a bright northern source  with $S_{870\,\mu\mathrm{m}}\approx25$\,mJy 
and a connecting bridge with $S_{870\,\mu\mathrm{m}}\approx7$\,mJy \citep{Miller}, as well as an offset satellite halo located 1.5\,Mpc from the core \citep{Hill20}.
The redshifts of two bright sources within the southern core were first constrained to lie at $4.300\pm0.002$ through  $^{12}$CO  lines from a blind ALMA 3-mm spectral scan  \citep{Strandet}. 
Deeper follow-up ALMA observations began in Cycles 3 and 4 and  initial results highlighted a core region of 14 SMGs \citep{Miller}. Recently the extended structure has been mapped by ALMA in Cycles 5 and 6 \citep{Hill20}.

This paper presents 
optical-through-near-IR photometry 
of SPT2349-56, with the aim of searching for additional optically-selected cluster members 
and to study the bright SMGs in the SPT2349-56 proto-cluster core. 
Section 2 describes the optical and near-IR data, while section 3 presents the analysis and results of our study. 
We discuss our results in section 4 and conclude in section 5.
A Hubble constant $H_0=70$\ km\,s$^{-1}$\,Mpc$^{-1}$ and density parameters $\Omega_{\Lambda}=0.7$ and $\Omega_{\rm m}=0.3$ are assumed throughout.

\begin{figure*}\centering
	\includegraphics[width=\textwidth]{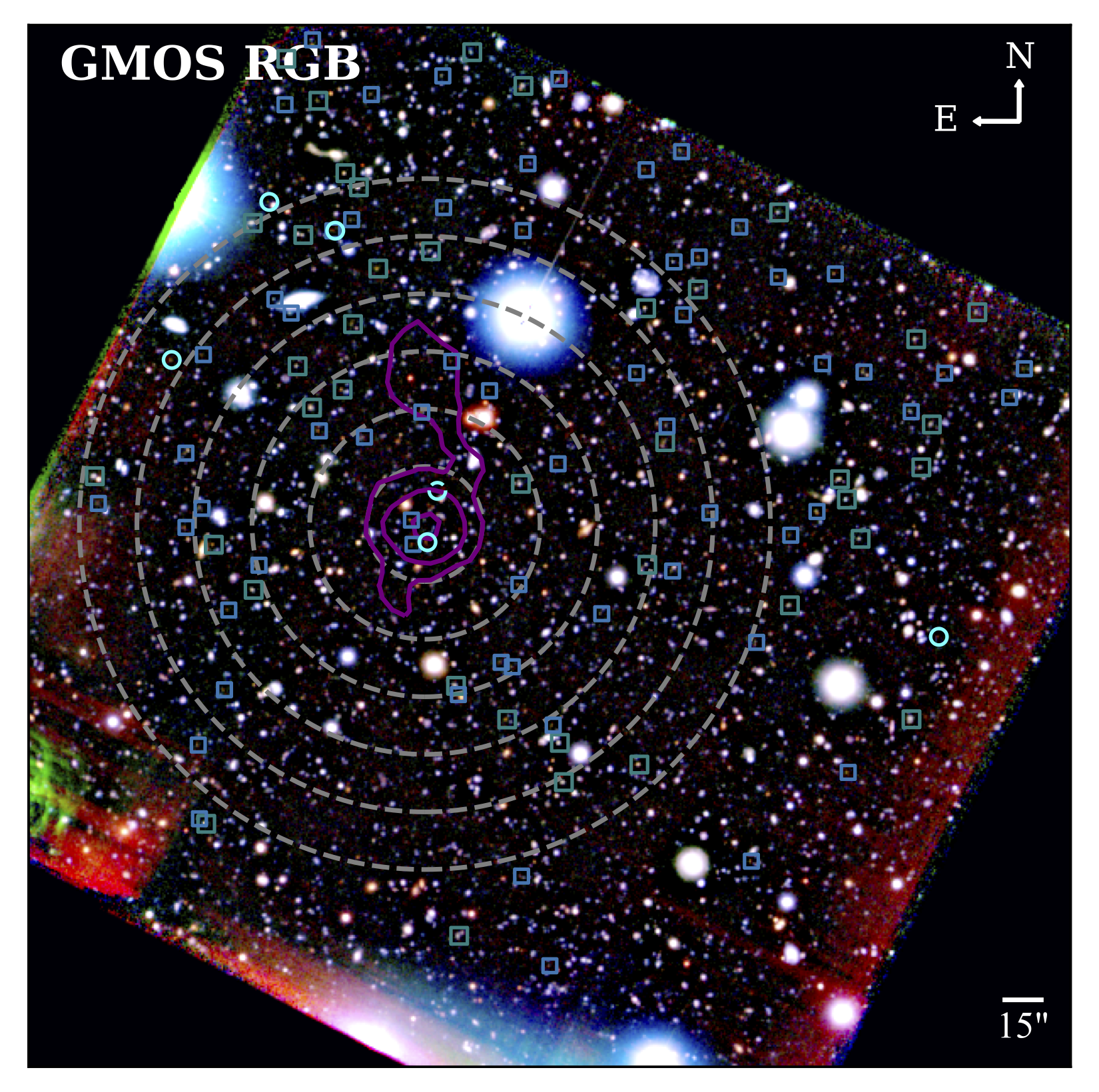}
	\caption[]{Gemini-S GMOS $g,r,$ and $i$-band RGB image ($\approx6\times6$\,arcmin$^2$) with LABOCA contours at $2\sigma$, $5\sigma$, and $10\sigma$ in purple. The $z\simeq4$ LBGs within the GMOS footprint selected using the \cite{Toshikawa18} colour criteria ($g-r>1,\ -1<r-i<1,\ {\rm and}\ 1.5(r-i)<g-r-0.8$) are identified by coloured squares. They are differentiated by $i$-band magnitude; ${>}15\sigma$ ($i\lesssim25$; large teal) and $5{-}15\sigma$ ($25\lesssim i\lesssim26.2$; small blue). 
	The cyan circles are additional LBG candidates that are selected when the $r-i$ colour criteria is relaxed to an upper limit of 2. The cyan dashed circle is LBG4, which was added upon a visual inspection of $g-r$ dropout candidates within the core region. The concentric dashed grey circles are at radial distances in increments of 0.35 arcmin from the centroid of ALMA sources, and are used for the radial analysis in Sect.~4.1.
	}
	\label{fig:2349lbg}
\end{figure*}

\section{Observations}

\subsection{Gemini-S imaging}
Imaging and spectroscopy of SPT2349-56 were obtained under programme ID GS-2017B-Q-7 (PI Chapman).
Deep Gemini imaging in the $g,r,i$ and $K_{\rm s}$ bands were obtained using GMOS (optical; \citealt{Hook}) and FLAMINGOS-2 (near-IR; \citealt{Eikenberry}) at the Gemini-South Observatory in Cerro-Pachon, Chile. The observations were performed in service mode under near photometric conditions on 2016 October 6 and 2016 November 23, with standard observing strategies. 
Data reduction followed Gemini-\textsc{iraf} reduction scripts and standard parameters for the optical data. The total integration times are $5{,}520,\, 5{,}160$, and $5{,}640$ sec and the seeing FWHM are 0.58, 0.58, and 0.60 for $g,r,$ and $i$-band images, respectively. The $g,r,$ and $i$-band fluxes were calibrated against DES imaging. Fig. \ref{fig:2349lbg} is an approximately $6\times6$\,arcmin$^2$ false colour RGB image of the GMOS $g,r,$ and $i$ bands - the SPT2349-56 core lies within the southern lobe of the structure highighted by LABOCA contours.

For the $K_{\rm s}$ observations, 
the data were reduced using the {\tt python}-based FLAMINGOS-2 Data Pipeline, \textsc{fatboy}, created by the Astronomy Department at the University of Florida. Briefly, a calibration dark was subtracted from the data set, a flatfield image and a bad pixel map were created, and the flatfield was divided through the data. Sky subtraction was performed to remove small-scale structure with a subsequent low-order correction for the large-scale structure. Finally, the data were aligned and stacked, resulting in a total integration time of $3{,}942$ sec. 
The mosaiced image was calibrated to the astrometry and photometry of 2MASS catalogues.
The seeing, as derived from the FWHM size of stars in each frame, ranged from 0.6 to 0.8\,arcsec. 

\textsc{SExtractor} \citep{Bertin} was used to extract catalogues of sources in all bands. The detection threshold and analysis threshold varied between bands from $1.1{-}2.5\sigma$ per pixel,  while the minimum detection area was kept constant at 3\,pixels. A standard Gaussian filter with a $5\times5$ convolution mask of a Gaussian PSF with a FWHM of 3\,pixels was applied to all images. \textsc{SExtractor} was used in single-image mode. Any offsets between  sources measured in different bands required cross-matching to within ${<}\,0.3$\,arcsec in GMOS and ${<}\,0.5$\,arcsec between GMOS and FLAMINGOS-2/IRAC. We verified visually that this did not result in any missed sources or erroneous associations. In the $g,r,$ and $i$ bands, the photometry was extracted within a 1.6\,arcsec aperture. $K_{\rm s}$-band photometry was extracted using $m_{\rm{auto}}$.
The $3\sigma$ AB magnitude depths achieved were $g=28.6,\  r=27.7,\  i=26.8,$
and $K_{\rm s}=24.5$.

\begin{figure*}\centering
	\includegraphics[width=\textwidth]{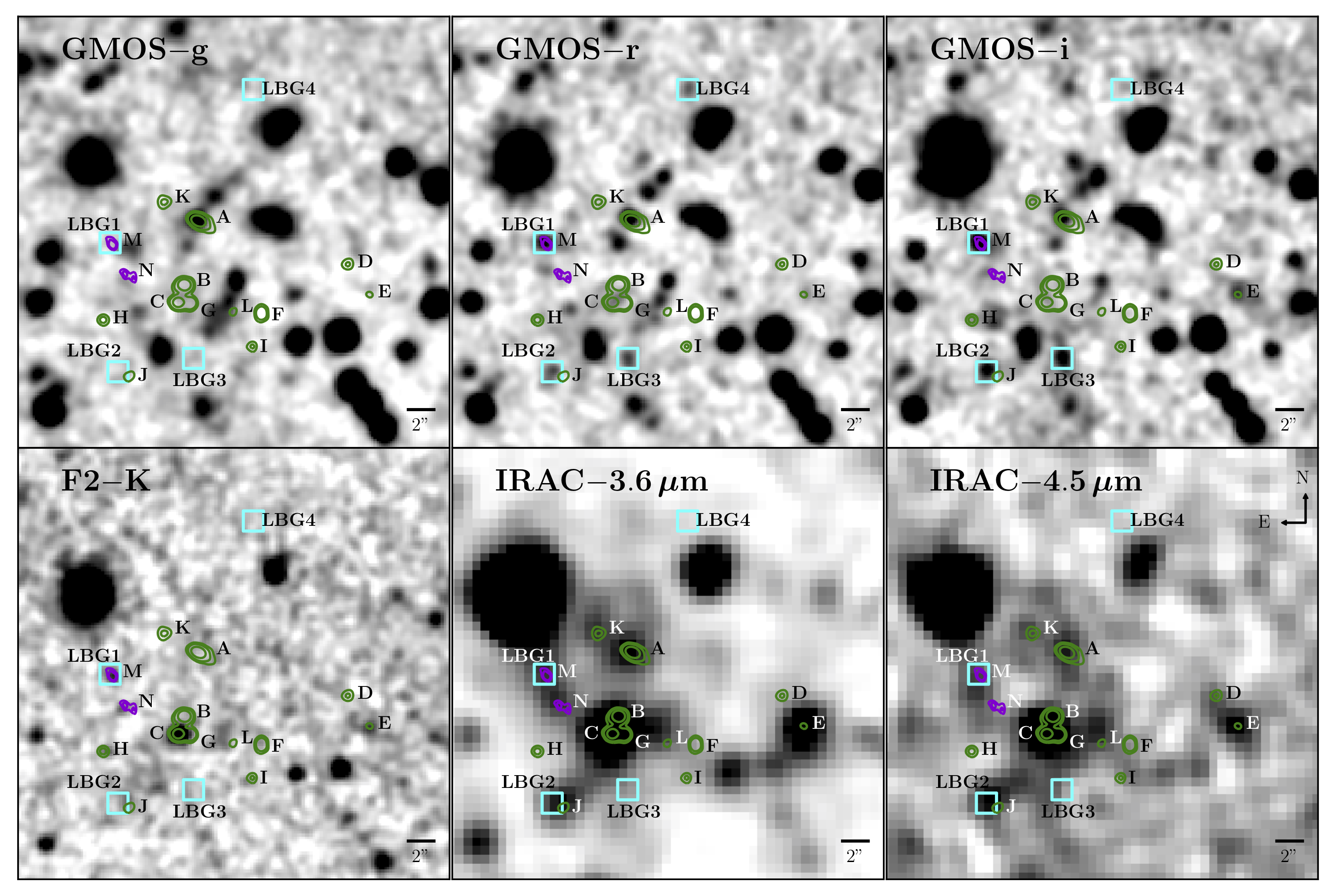}
	\caption[Observed frame optical through mid-IR  and imaging for SPT2349]{Optical and near-IR imaging ($30\times30$\,arcsec$^2$ cutouts). {\it Top row}: Gemini-S GMOS $g,r,$ and $i$-band ($0.47$, $0.63$, $0.84$\,$\mu$m); {\it Bottom row}: Gemini-S FLAMINGOS-2 $K_{\rm s}$ band ($2.16\,\mu$m), \Spitzer IRAC $3.6\,\mu$m, and \Spitzer IRAC $4.5\,\mu$m. The green contours are drawn at $5\sigma$, $10\sigma$, and $30\sigma$ and show the ALMA 358-GHz continuum data. The purple contours are drawn at $3\sigma$ and $4\sigma$ and show the average ALMA Band 7 channel map covering the \CII\ emission in sources $M$ and $N$. The  $z\simeq4$ LBGs identified in this work (`LBG1--4') are shown with cyan squares. 
	}
	\label{fig:2349opt}
\end{figure*}

\subsection{IRAC imaging}
The SPT2349-56 field was twice observed at 3.6 and 4.5\,$\mu$m with the Infrared Array Camera (IRAC; \citealt{Fazio04}) on board the {\sl Spitzer Space Telescope} \citep{Werner04}.  It was first observed in 2009 August as part of a large programme to obtain follow-up imaging of a large sample of SPT-selected SMG sources (PID\,60194; PI Vieira). The observing scheme used was to obtain 36 dithered 100-sec integrations at 3.6\,$\mu$m and, separately, a much shallower $12\times30$-sec integration at 4.5\,$\mu$m.  Later, in Cycle 8, the field was covered serendipitously as part of the {\it Spitzer\/}-SPT Deep Field survey (PID\,80032; PI Stanford, \citealt{Ashby13}).  This surveyed 92\,deg$^2$ uniformly in both IRAC passbands with an integration of $4\times30$\,sec.  Using established techniques, we combined all exposures covering the SPT target from PID\,60194 and 80032 at 3.6 and 4.5\,$\mu$m to obtain the best possible S/N in our final mosaics, which were pixellated to 0.6\,arcsec. 
The maps were shown in \cite{Miller} to illustrate possible identifications of ALMA sources, but the IRAC photometry was not extracted or tabulated. 
Here we compile faint (${\approx}\,3\sigma$) catalogues by running \textsc{SExtractor} on the combined maps. A variety of \textsc{SExtractor} parameters (primarily the analysis and detection threshold) were adjusted to optimize detection of the well-separated sources. We did not attempt a deconvolution of the crowded field.
%
%
Eight of the 14 sources identified by ALMA are detected in the IRAC bands at ${>}\,3\sigma$ in at least one of the 3.6 or 4.5\,$\mu$m channels, as shown in Fig. \ref{fig:2349opt}. These are listed with 4\,arcsec-aperture magnitudes (corrected to total flux using point spread function curve of growth) in Table \ref{table:dataSMGphot}.

The brightest SMGs in the core are labelled $A$ through $N$, as ordered by their 850\,$\mu$m flux density \citep{Miller}. 
The central region of the proto-cluster is marked by three (possibly interacting) luminous ALMA sources ($B,C,$ and $G$) spanning a 2\,arcsec-diameter region. These three galaxies are highly confused in the IRAC bands, but are clearly dominated by a bright, possibly extended source whose centroid lies close to source $C$. The $K_{\rm s}$-band data, with ${\approx}\,0.6$\,arcsec seeing, easily resolves this trio, but only significantly detects $C$ as a $K_{\mathrm{AB}}=22.2$ isolated source. 
The $K_{\rm s}$-band limit of sources $B$ and $G$ ($K_{\rm s}>24.5$) allows us to 
place limits on the IRAC flux that $B$ and $G$ may contribute. For those ALMA sources detected at both bands, the colour range is $K_{\rm s}-3.6\,\mu\rm{m}=1.2{-}2.3$. Sources $B$ and $G$ may therefore be as bright as $22.2{-}23.3$ in IRAC 3.6\,$\mu$m.
This range represents a negligible contribution to the IRAC flux of source $C$, and corresponds to about 3$\times$RMS to 1$\times$RMS of the IRAC data in uncrowded regions of the map. We therefore  adopt this same 3$\times$RMS limit as for the other IRAC-undetected sources (listed in Table \ref{table:dataSMGphot}).

\begin{table*}
\caption[dum]{\small Optical and Near-IR Photometry of SPT2349-56 SMGs.}
\label{table:dataSMGphot}\begin{threeparttable}
\centering
\begin{tabular}{lcccccc}\hline
ID & $g$ & $r$ & $i$ & $K_{\rm s}$ & $3.6\,\mu$m & $4.5\,\mu$m \\
\hline
{} & [AB] & [AB] &   [AB] &  [AB] &  [AB] &  [AB]  \\
\hline
A & 26.40$\pm$0.02 & 25.93$\pm$0.02 & 25.77$\pm$0.03 & -- & 21.93$\pm$0.02 & 21.84$\pm$0.02 \\
B & -- & -- & --  & -- & -- & -- \\
C & -- & 27.06$\pm$0.05 & 27.33$\pm$0.10 & 22.21$\pm$0.08 & 20.83$\pm$0.01 & 20.79$\pm$0.01 \\
D & -- & -- & -- & -- & -- &  22.82$\pm$0.22 \\
E & -- & 28.13$\pm$0.14 & 26.28$\pm$0.04 & 24.03$\pm$0.12 & 21.73$\pm$0.02 & 21.68$\pm$0.01 \\
F & -- & -- & -- & -- & -- & -- \\
G & -- & -- & -- & -- & -- & -- \\
H & -- & -- & -- & 24.31$\pm$0.15 & -- & -- \\
I & -- & -- & -- & -- & -- & -- \\
J & -- & -- & -- & 23.54$\pm$0.10 & 22.23$\pm$0.02 & 22.12$\pm$0.02  \\
K & -- & -- & --  & -- & 22.78$\pm$0.08 & 22.67$\pm$0.24 \\
L & -- & -- & -- & -- & -- & -- \\
M & -- & 26.35$\pm$0.03 & 25.54$\pm$0.02 &  23.82$\pm$0.11 & 22.67$\pm$0.08 & 22.10$\pm$0.13 \\
N & -- & -- & --  & 24.29$\pm$0.15 & 22.68$\pm$0.08 & -- \\
\hline
\end{tabular}
\begin{tablenotes}
\item Note: For sources without entries the $3\sigma$ limits are $g=28.6,\  r=27.7,\  i=26.8,\  K_{\rm s}=24.5,\  3.6\,\mu\rm{m}=22.1,\ {\rm and}\ 4.5\,\mu\rm{m}=22.1$. 
\end{tablenotes}\end{threeparttable}
\end{table*}

\subsection{VLT spectroscopy}

We also observed SPT2349-56 with the X-shooter echelon spectrograph \citep{Xshooter} on the ESO Very Large Telescope (VLT)-UT2, Kueyen, as part of programme 092.A-0503(A) (PI Chapman). X-shooter is capable of near-continuous spectroscopy from 0.3 to $2.48\,\mu$m, with a slit width and length of 1.2 and 11\,arcsec, respectively.  
We observed two positions centred on the optical/near-IR identifications of ALMA components $A$ and $C$, dithering the observations in an ABBA sequence at positions $+3$\,arcsec and $-3$\,arcsec along the slit axis every 600\,sec. 
We first peaked up on a nearby star in a field within 1\,arcmin of the target position, then did a blind offset.

Observations were taken on the nights of UT 2013 October 16 and November 12, with total integrations of $5{,}400$\,sec for each source. 
Seeing conditions were similar throughout these observations with values around 0.8\,arcsec and taken with a low average airmass of 1.2.

We used the ESO pipeline \citep{Modigliani} to reduce our data.  This pipeline applies spatial and spectral rectification to the spectra 
using the two-dimensional arc spectra. 
The data were flatfielded and cosmic rays were identified and masked.  The two dither positions were subtracted to remove the sky to first order, and the different echelle orders were combined together into a continuous spectrum. Flux calibration was achieved through observations of standard stars LTT3218, GD-71 and Feige 110.

\subsection{ALMA observations}

Deep ALMA observations covering SPT2349-56 were presented in \cite{Miller} and \cite{Hill20}. In this work we make use of the Band 7 maps covering the redshifted \CII$_{\rm 158\,\mu m}$ fine structure line ($\nu_{\rm obs}=358.4$\,GHz at the median cluster redshift). Data reductions and processing are described in these works. The maps reach an average depth of 0.1\,mJy RMS at this frequency, corresponding to a 3$\sigma$ limit on the star formation rate (SFR ${<}\,10$\,\msunrate).

\section{Results}

\subsection{Rest-frame ultraviolet properties of the SPT2349-56 core}
We first measure the $g,r,$ and $i$-band properties of SPT2349-56 galaxies, which are at rest-frame ultraviolet wavelengths at $z=4.3$.
Four ALMA sources ($A,C,E,$ and $M$) appear to have significant counterparts in the Gemini-S optical imaging (see Fig.\,\ref{fig:2349opt} and Table \ref{table:dataSMGphot}). However, one of the identifications is  
offset from its respective ALMA source: $A$ (0.4\,arcsec). 
This was assessed by aligning the optical images to the $K_{\rm s}$ and IRAC astrometric frames using many bright stars in the field. The $K_{\rm s}$ and IRAC images were aligned to the 2MASS astrometric frame, which was verified to provide a good match to the ALMA frame -- several near-IR identifications of ALMA sources are all well centred (specifically, $C,E,K,$ and $M$ all show agreement within 0.2\,arcsec with the ALMA centroids). Three of the ALMA sources detected in the optical drop out in the $g$ band, consistent with $z=4.3$ galaxies. The optical identification for source $A$, however, is bright in the $g$ band; while the ALMA source $A$ is confirmed to be at $z=4.3$, the optical source was identified spectroscopically to be a foreground $z=2.54$ galaxy, as described in Sect.~\ref{sec:A}.

We also use the optical imaging to search for LBGs using the $g$-band dropout technique of \cite{Toshikawa18}, specifically $g-r>1,\ -1<r-i<1,\ {\rm and}\ 1.5(r-i)<g-r-0.8$. These colour criteria are sensitive to galaxies in the $z=3.3{-}4.2$ range. We require a ${>}5\sigma$ and ${>}3\sigma$ detection in the $i$ and $r$ bands, respectively. LBG candidates are identified with coloured squares in Fig. \ref{fig:2349lbg}. LBG surveys set a tight $r-i$ colour selection window to reduce the number of low-$z$ interlopers. This compromise ensures a pure sample of similar-redshift LBGs but is at the expense of excluding some bonafide high-$z$ member galaxies \cite[e.g.][]{Steidel99}. We noticed several potential LBGs we identified visually as clear $g-$band dropouts were selected by relaxing the upper limit of the $r-i$ colour to 2. This strongly suggests we have highly reddened LBGs in this dense environment, a  finding that is consistent with the overdensity of bright SMGs, which implies additional fainter, dust-obscured galaxies may be present as well. We have therefore also applied a selection with a relaxed upper limit of the $r-i$ colour to 2. 
However, we note that we are incomplete (potentially missing some sources) for $i_{\rm AB}>25.7$ when doing so. These LBG candidates are identified with cyan circles 
in Fig. \ref{fig:2349lbg}. The colour-selection windows are illustrated in Appendix~B.
While we focus on the core ALMA region here, we also present the LBG population in the wider field in Appendix~B and discuss the results in Sect.~4.1. Additionally, we searched the core region for sources with large $g-r$ colour (irrespective of the $i$-band magnitude) and found a further  candidate with a strong $g-r$ dropout that our search did not identify as its $i$-band magnitude falls below the 5$\sigma$ detection threshold, with a very blue $r-i<-1$.
This source has a spectroscopic confirmation from its strong Ly$\alpha$ emission in Apostolovski (in prep.), likely driving the blue $r-i$ colour. ALMA sources $C$ and $E$ are also undetected in the $g$ band, however, their $r-i$ colours are either too faint or too red to satisfy the LBG criteria above. This results in 
four LBGs 
identified within this core structure (labelled LBG1--4), 
two of which lie near ALMA sources $J$ and $M$. 

\begin{figure}\centering
	\includegraphics[width=8.2cm]{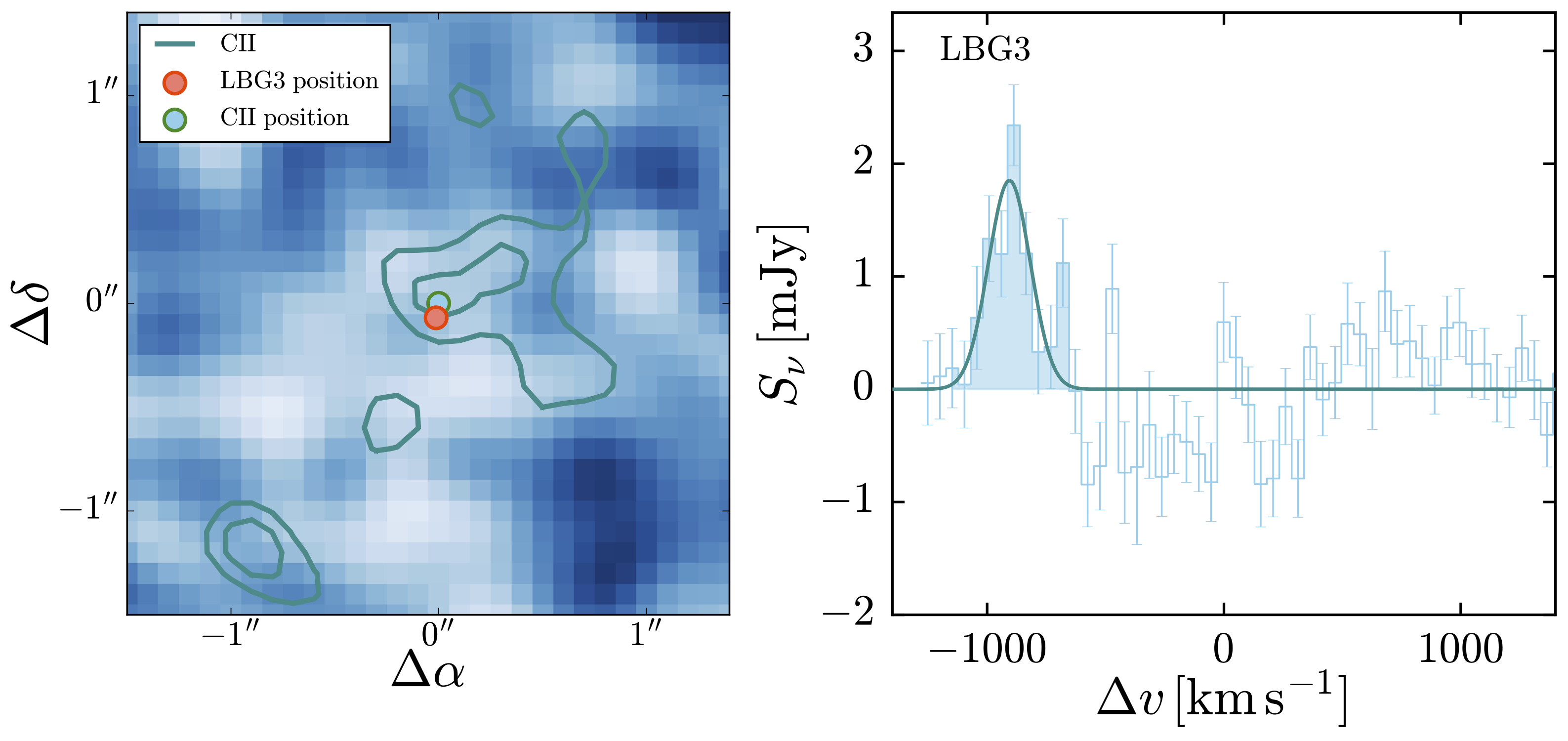}
	\caption[C+ in LBG3]{ALMA spectrum of LBG3 highlighting a $6.8\sigma$ detection of \CII\ at a redshift $z=4.288$. This source was not previously catalogued in the work of \cite{Hill20}. {\it Left:} Continuum image with \CII\ contours overlaid  at $2\sigma$, $3\sigma$, and $4\sigma$. {\it Right:} The 1D spectrum, with zero indicating no velocity relative to the mean redshift of the proto-cluster core at $z=4.303$. 
	}
	\label{fig:2349lbg3}
\end{figure}

While LBG1 lies 0.2\,arcsec from the ALMA centroid of source $M$ and is likely the same galaxy, LBG2 and the ALMA centroid for source $J$ are offset by about 0.8\,arcsec (see Fig.\,\ref{fig:2349opt}) and we therefore treat LBG2 as a separate galaxy in this work. Deep spectroscopic follow-up could ascertain whether or not this second LBG is close in velocity to ALMA source $J$.
The third and fourth LBGs in the core region lie 5\,arcsec south and 15\,arcsec north (respectively) of the bright central source $C$. 
There is no (sub-)mm continuum detected at these positions in the deep ALMA maps reaching $S_{850\,\mu\text{m}}<0.3$\,mJy, 3$\sigma$ \citep{Hill20}. 
For LBG3 there is a candidate $6.8\sigma$ \CII$_{158\,\mu\text{m}}$ line detection at 359.4\,GHz ($z=4.288$), a significance that places it just below the cutoff adopted in the catalogue of \cite{Hill20}. 
It has a line flux of  
$0.46\pm0.07$\,Jy\,km\,s$^{-1}$ and a FWHM of $204\pm35$\,km\,s$^{-1}$. 
The rest-UV SFR estimate (calculated as described in Table\,\ref{table:dataLBG}) is 
20\,\msunrate, which agrees reasonably well with the \CII\ line strength for typical $z\sim4$ galaxies \citep{Schaerer}. The \CII\ emission is shown in Fig.\,\ref{fig:2349lbg3}.
In contrast, LBG4 shows no evidence of a \CII\ line. However, it is identified as the brightest Ly$\alpha$ emitter in the MUSE survey of Apostolovski (in prep.) and is clearly a member of the proto-cluster, with a Ly$\alpha$ redshift of 4.308.
Neither of these LBGs are detected at $K_{\rm s}$ or IRAC wavelengths. 
As demonstrated, relaxing the colour criteria to include fainter and redder LBGs within the proto-cluster core did not contaminate the sample with line-of-sight, low-$z$ sources.

\begin{table}
\setlength{\tabcolsep}{5pt}
\caption[dum]{\small Properties of SPT2349-56 LBGs. }
\label{table:dataLBG}\begin{threeparttable}
\centering
\begin{tabular}{lccccc}\hline
ID &  RA & Dec & $r$ & $i$ &  SFR \\
\hline
 &  &   &  [AB] &  [AB]   & [\msunrate] \\
\hline
1$^{\rm{a}}$ &  23:49:43.406 & $-$56:38:20.93  & 26.35 & 25.54 & 24$\pm$4 \\ 
2$^{\rm{b}}$ &  23:49:43.340 & $-$56:38:29.90 & 26.83 & 25.86 & 18$\pm$6 \\ 
3 &  23:49:42.703 & $-$56:38:28.97 & 26.77 & 25.76 & 20$\pm$6 \\ 
4$^{\rm{c}}$ & 23:49:42.198 & $-$56:38:10.28 & 27.07 & 28.11 & 4.5$\pm$1 \\ 
\hline
\end{tabular}
\begin{tablenotes}
\item Note: Photometry errors range from 0.02 for the brightest detections to 0.2 for the faintest, as listed in Appendix~B. 
None of these four LBGs are significantly detected in the $g$ band. SFRs for LBG1--3 are calculated as $1.4\times10^{-28}$ L$_\nu$($1,500$\,\AA) \msunrate, with extinction estimated from $r-i$ colour. The SFR for LBG4 is estimated from its Ly$\alpha$ emission.
\item[a] This LBG is well aligned with ALMA source $M$ and we treat it as such above. We duplicate its properties here for completeness.
\item[b] This LBG lies 0.8\,arcsec offset from ALMA source $J$. We treat this LBG as a distinct galaxy since the $K_{\rm s}$ and IRAC fluxes are well aligned with source $J$.
\item[c] LBG4 is identified as the brightest $z=4.30$ Ly$\alpha$ emitter in Apostolovski (in prep.). 
\end{tablenotes}\end{threeparttable}
\end{table}

\subsubsection{Foreground source of SMG $A$}\label{sec:A}
\begin{figure}\centering
	\includegraphics[width=8.2cm]{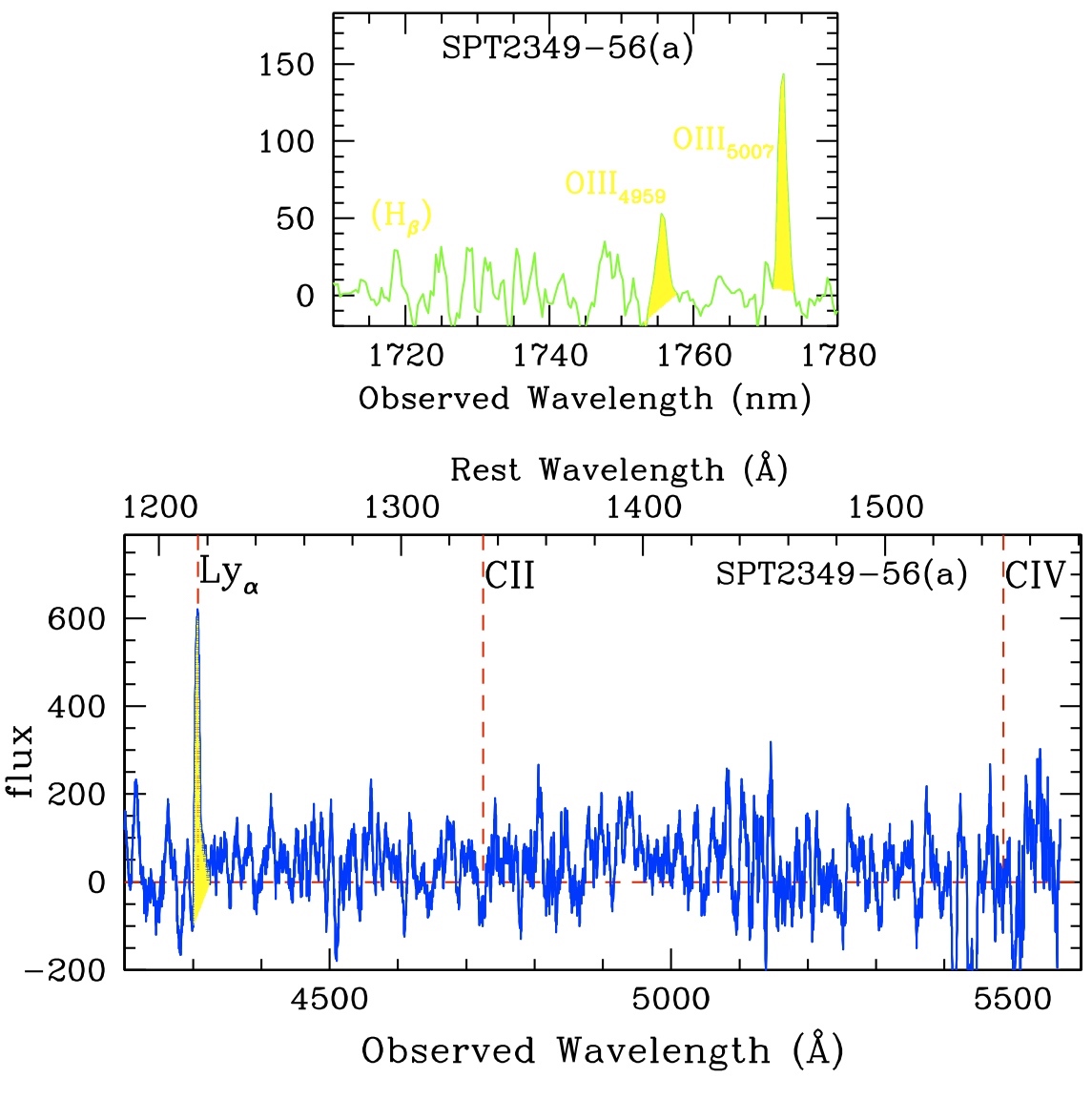}
	\caption[X-shooter spectrum of source $A$]{The VLT/X-shooter spectrum of source $A$. The spectrum does not show any $z=4.3$ features at any wavelength. However, it does reveal a foreground $z=2.54$ star-forming galaxy that likely dominates the  $g,r,$ and $i$-band photometry. The {\it top panel} shows a cutout from the near-IR arm of X-shooter with the $z=2.54$ \OIII\ lines detected. The {\it bottom panel} shows the corresponding UV-arm spectrum, detecting Ly$\alpha$ at the same $z=2.54$. The faint $r=25.9$ source has a linewidth of 53\,km\,s$^{-1}$ and is unlikely to be very massive -- it does not provide any significant gravitational lensing boost to source $A$ (which we estimate as $\mu<1.2$).}
	\label{fig:2349xshooter}
\end{figure}

Source $A$ stands out in the rest-UV images ($g,r,$ and $i$ bands), as it is brighter than any other SPT2349-56 source and it is the only one detected in the $g$ band. An X-shooter spectrum of source $A$ (see Fig.\,\ref{fig:2349xshooter}) does not show any emission features expected from a $z=4.3$ galaxy, but does reveal a foreground $z=2.54$, star-forming galaxy that likely dominates the $g$-band photometry and contributes to the $r$ and $i$-band fluxes. The optical band centroids are significantly offset by 0.4\,arcsec from the ALMA and IRAC emission centroids.  
The foreground source is unlikely to be very massive given the blue colours, an $r$-band magnitude of $\approx26$, and an \OIII$_{\,5,007\,\text{\AA}}$ line FWHM of 53\,km s$^{-1}$. The linewidth suggests an upper limit to a dynamical mass enclosed within a 2\,kpc-radius of $M_{\rm dyn}=1.56\times R_{1/2}\ \sigma^2<1.6\times10^{9}$\,\msun\ \citep[see][]{Erb06}.
For $z\simeq2.5$ LBGs of this luminosity, typical stellar masses of $<10^{9}$\,\msun\ are in agreement with our dynamical estimate \citep{Shapley05}. It is difficult to directly ascertain the stellar mass of this galaxy, but it is clearly undetected in the $K_{\rm s}$ band down to 24.5 magnitudes. Due to the offset relative to the IRAC source we ascribe the faint IRAC flux to the $z=4.3$ ALMA source $A$ and not to the foreground UV-luminous galaxy. 

This configuration does not provide a significant gravitational lensing boost to source $A$, 
even assuming a lensing mass 
reflective of the high end of our dynamical mass estimate. 
Using a simple lens model \citep{Spilker} we set the Einstein lensing mass to $2.5\times M^*$ with $M^*=1.6\times10^{9}$\,\msun\ (a generous assumption) and adjust the background source offset until the apparent image is 0.5\,arcsec from the lens position to match our configuration. We find a magnification factor of $\mu=1.15$ for a circular lens, or ranging from 1.09 to 1.24 for a highly elliptical lens ($e=0.6$), depending on the position angle. 

\subsection{Rest-frame optical properties of the SPT2349-56 core}

\begin{table}
\caption[dum]{\small Properties of SPT2349-56 SMGs.}
\label{table:dataSMGprop}\begin{threeparttable}
\centering
\begin{tabular}{lccccc}\hline
ID & A$_V$ & Age & $\tau_{\rm SFH}$ $^{\rm{b}}$ & $\log{\left(\sfrac{M^*}{\rm{M}_{\odot}}\right)}$ & $f_{\rm gas}$ $^{\rm{c}}$ \\
\hline
{} & {[mag]} & [Gyrs] & [Gyrs] & {} & {} \\
\hline
A & 2.3$\pm$1.1 & 1.0$\pm$0.4 & 1.6$\pm$2.3 & 11.35$\pm$0.30 & 0.35 \\
B & -- & -- & --  & $<$10.76 & $>$0.66 \\
C & 1.3$\pm$0.1 & 1.2$\pm$0.3 & 1.6$\pm$1.9 & 11.51$\pm$0.24 & 0.17 \\
D$^{\rm{a}}$ & -- & -- & -- & 10.89$\pm$0.29 & 0.52 \\
E & 2.2$\pm$0.7 & 0.8$\pm$0.5 & 2.5$\pm$2.5 & 11.15$\pm$0.28 & 0.25 \\
F & -- & -- & -- & $<$10.76 & $>$0.37 \\
G & -- & -- & -- & $<$10.76 & $>$0.34 \\
H$^{\rm{a}}$ & -- & -- & -- & 10.89$\pm$0.35 & 0.36 \\
I & -- & -- & -- & $<$10.76 & $>$0.28 \\
J & 0.9$\pm$0.7 & 1.1$\pm$0.4 & 1.0$\pm$1.9 & 11.04$\pm$0.19 & 0.17 \\
K & 1.6$\pm$1.1 & 1.0$\pm$0.4 & 1.5$\pm$2.2 & 10.88$\pm$0.27 & 0.29 \\
L & -- & -- & -- & $<$10.76 & $>$0.37 \\
M & 1.0$\pm$0.5 & 0.9$\pm$0.4 & 2.4$\pm$2.5 & 10.57$\pm$0.23 & 0.24 \\
N$^{\rm{a}}$ & -- & -- & -- & 11.17$\pm$0.24 & 0.06 \\
\hline
\end{tabular}
\begin{tablenotes}
\item Note: The first four columns are best-fitting parameters from \textsc{cigale}.
\item[a] \textsc{cigale} fits are performed on these sources with the age of the main population fixed, as photometry is faint and sparse in wavelength coverage. 
\item[b] Exponentially declining star formation histories, $\rm{SFH}\propto e^{-t/\tau}$
\item[c] Gas fraction, $f_{\rm gas}=\sfrac{M_{\rm gas}}{\left(M_{\rm gas}+M^*\right)}$; gas masses were determined from the $^{12}$CO line luminosities (\CII\ when $^{12}$CO was not available) from \cite{Miller} using the standard conversion of \cite{Bothwell} with $\alpha_{\rm CO}=0.8$ and a $1.36\times$ correction factor to include helium.
\end{tablenotes}\end{threeparttable}
\end{table}

We consider next the observed $K_{\rm s}$, IRAC $3.6\,\mu$m, and IRAC $4.5\,\mu$m properties of the SPT2349-56 sources with an aim of constraining stellar mass.
As noted in Sect. 2, we detect eight sources in one or both IRAC bands (namely $A,C,D,E,J,K,M,$ and $N$). Five of these are detected in $K_{\rm s}$ ($C,E,J,M,$ and $N$), three of which are also detected in $i$ and $r$ ($C,E,$ and $M$). 
In addition, source $H$ is detected in just the $K_{\rm s}$ band.
Such generally incomplete or low SNR photometry limits the constraints possible from spectral energy distribution (SED) fitting (the exceptional source $C$ is discussed separately below).

We estimate the 
stellar masses by modelling the multi-wavelength SEDs using the software {\it Code Investigating GALaxy Emission} (\textsc{cigale}; \citealt{Noll09,Serra11,Boquien19}). 
\textsc{cigale} adopts an energy balance principle between the UV-optical and far-IR-mm regimes -- the energy absorbed by dust in the UV-optical is proportional to the thermal radiation emitted by dust in the far-IR. 

We first consider the six sources detected in both IRAC bands, which also generally have supporting detections in other bands. To model the sources we have assumed a delayed star-formation history with a single exponential decrease. The e-folding time ($\tau$) and age of the stellar population are kept as free parameters while a solar metallicity and a Chabrier initial mass function (IMF; \citealt{Chabrier}) are assumed. 
It has been found in the literature \citep{Michalowski12} that stellar mass estimates using a single-exponential decay SFH can be a factor of about 2 different than double-exponential or a bursty-type SFH. Typically the single exponential decay SFH provides the lowest stellar mass, our estimates are therefore relatively conservative.
Nebular emission is included in the fitting using templates from \cite{Inoue11}. The dust attenuation is modelled using a \cite{Calzetti00} attenuation curve with a power law slope of zero. The stellar mass estimates are consistent within the errors even when the slope is allowed to vary. 

AGN contribution can affect the stellar mass estimate depending on the AGN type and the available photometry. A detailed study of the effects of AGN contribution on estimates of various physical properties including stellar mass when using \textsc{cigale} was carried out by \cite{Ciesla15}. To summarize, depending on the type of AGN present, the stellar mass can be under- or over-estimated on average by 20 per cent when using similar bands as the ones we have available 
\citep[refer to Fig. 11 of][]{Ciesla15}. Type-I AGN can contribute heavily in the rest-frame UV-optical as well as mid-IR. However, the extinction values for the SMGs in SPT2349-56 rule out large contribution to the luminosity by a type-I AGN. 

We do not have sensitive photometry in the mid-IR to constrain contribution from a type-II AGN, which would primarily contribute to the rest-frame mid-IR \citep[refer to Fig. 4 of][]{Ciesla15}.
For source $C$, which does have good coverage in rest-frame optical, near-IR and far-IR, we estimated the stellar mass by fixing the AGN contribution to a type-II template, ranging from 0.0 to 0.9. 
As is the case for source $C$, the remaining sources show no evidence for a type-I AGN. Additionally, 
the dust peak in the far-IR for the remaining sources is also unconstrained with our (sub)-mm photometry, and the SED fitting 
was primarily conducted across the optical/near-IR region. In this case, zero contribution was assumed from an AGN component.

Importantly, \textsc{cigale} can take upper limits as inputs in multiple bands and incorporate them in the reduced $\chi^2$ calculation \citep{Boquien19}. For non-detections we set the limits to our $3\sigma$ magnitude depths. Even though they are non-detections, they contribute to the reduced $\chi^2$ value, the shape of the SED, as well as inferred physical parameters. For optical peaks that fall just below the $3\sigma$ limit, we continue to use the extracted photometry, treating it as a detection in the SED fitting, as it places more stringent requirements on the fitting. Apart from source $C$, we note that the sources are slightly over-fitted due to their low number of detections, which is indicated by their reduced $\chi^2$ values of less than 1 (see Appendix~A). Given the currently available data, however, we deemed this the most sensible solution. (In the modelling of source $A$, the $g,r,$ and $i$-band photometry attributed to the foreground galaxy are treated as upper limits.)

The stellar masses for the sources are estimated using the `pdf analysis' module in \textsc{cigale}. Mock catalogues are generated and analyzed to check the reliability of these estimates within the parameter space explored. 
Each discrete model corresponds to a reduced $\chi^2$ value \citep{Boquien19}. The array of reduced $\chi^2$ values for each model is then used to create a probability distribution for the stellar mass. This distribution is generally not a Gaussian and a value marginalized over the distribution is calculated. The standard deviation of this probability distribution is quoted as the error on the stellar mass. We note that the stellar mass estimated from the probability distribution does not necessarily correspond to the best-fitting model. The best-fitting SED only corresponds to the model with the smallest reduced $\chi^2$ value, i.e. only the peak of the distribution. Based on the best-fitting model for a source, \textsc{cigale} generates a mock catalogue by modifying each quantity in the best-fitting model. This modification is done by adding a value taken from a Gaussian distribution with the same standard deviation as the original quantity. The catalogue is then fit using the same parameter space as the original and physical parameters are estimated again. If the mock results and original results show consistent stellar mass estimates, we consider the results reliable within the parameter space.

We find the stellar mass values reported here to be consistent with mock catalogue results and are thus reliable within errors quoted. Furthermore, the probability distributions for all sources do not show any significant bi-modality confirming the mock results.
Additionally, we found stellar mass estimates to be broadly consistent between fitting with and without ALMA data (adopted from \citealt{Miller}), where deviations between the two are less than one tenth of the error of the estimates on a log scale.
As discussed by \citep{Ciesla15}, fitting using only optical/near-IR photometry can give rise to 
systematic uncertainties in estimations of various physical properties -- up to about 20 per cent for stellar mass estimates. Results are shown in Appendix~A. 

We next estimate the stellar masses for those sources with more limited photometry: $D, H$ and $N$. Here we constrain the SED fits, allowing the age of the stellar population to lie between $0.7{-}1.25$ Gyrs, consistent with that of the brighter sources in the proto-cluster. We make the assumption that all galaxies in the core have a similar age due to the dense environment. While we have no clear way of confirming this, we also do not have any evidence to the contrary and deem a consistent age to be the most appropriate assumption given they live in the same halo now.
For sources $H$ and $N$ the results of the mock catalogue analysis are consistent within error of the best fit.
However, mock catalogue analysis for source $D$ showed the stellar mass estimate to not be as certain as for the others (as can be expected for fitting with a single-band detection) -- the limits at our other wavelengths did not provide strong enough constraints.
Best-fitting \textsc{cigale} results for these three sources are also shown in Appendix~A. 

The remaining five sources are undetected in all observed bands, 
and may actually have very low stellar masses. 
However, their large $^{12}$CO luminosities {\citep{Miller}} suggest sizable gas masses, often comparable to the other sources, and correlating reasonably with their SFRs. 
Their large $^{12}$CO line widths {\citep{Miller}} also suggest sizable dynamical masses. Very high gas fraction galaxies are possible, since at such an early epoch it is reasonable that this star formation episode represents the first major stellar growth phase in these galaxies, stimulated by the dense environment \citep{Rennehan}. 
However, we do not discount the possibility that some of these five SMGs have extreme dust extinction levels, as seen in some field SMGs \citep[e.g.][]{Simpson15} and that their stellar masses are sizeable, implying similar gas fractions to the other SMGs.

For the nine sources constrained by SED fits, 
we determine the stellar mass to lie in the range 
$(0.4{-}3.2)\times10^{11}$\,\msun, with a median \mstar\ of $(1.1\pm0.8)\times10^{11}$\,\msun,
where the error is the standard deviation. 
Stellar mass estimates are presented in Table~\ref{table:dataSMGprop}. Source $C$ is discussed in detail below. The cumulative stellar mass for the SPT2349-56 core is 
$(12.2\pm2.8)\times10^{11}$\,\msun, which is a lower limit as we have only included nine of the 14 sources.

For the remaining five sources, which lack 
any optical/near-IR detections, we estimate an upper limit on \mstar\ as below.
The majority of the stellar mass in a galaxy manifests itself as a rest-frame near-IR bump in the SED whose emission peaks at approximately $1.6\,\mu$m. For SPT2349-56 at $z=4.3$, this peak in stellar light is redshifted to $8.5\,\mu$m. Our closest observed photometry comes from IRAC's 4.5-$\mu$m band. We use the average mass-to-light ratio for the nine 
detected sources, \mbox{$\langle M^*/L_{4.5\,\mu\rm{m}} \rangle=1.09$\,\msun/\lsun},
to constrain the stellar mass in those five sources not robustly detected, shown as upper limits in Table~\ref{table:dataSMGprop}.

We also list the gas fractions (calculated as described in the notes of Table~\ref{table:dataSMGprop}), showing an average $f_{\rm gas}=0.3$ for the nine near-IR detected sources, and a limit of 
$f_{\rm gas}>0.3$ for the 5 undetected.  Thus, the estimates of the \mstar\ limits above are in reasonable agreement with the \mstar\ values one would infer from their $M_{\rm gas}$, even if extreme extinction is responsible for their non-detections. (See also the discussion in Sect.~4.3 and Fig.~\ref{fig:masses} regarding the relative masses of each source).

\subsubsection{A large stellar mass for source $C$}
As noted, source $C$ stands out with an exceptionally large rest-optical luminosity for $z=4.3$, especially compared to all other ALMA sources in the proto-cluster core ($\approx2.5$ times brighter than the next most luminous sources, $E$ and $A$). It lies near the centre-of-mass of the structure, and is embedded in a dense region of ALMA sources, possibly the core of a forming brightest cluster galaxy (BCG).
The X-shooter spectrum of the source detects a very faint continuum through the $H$ and $K_{\rm s}$ bands, but does not detect any emission lines. Specifically, at the wavelengths of the redshifted \OII$_{\,3,727\,\text{\AA}}$ and \MgII$_{\,2,800\,\text{\AA}}$ lines, there is no obvious excess. However, the observed wavelengths do lie within relatively noisy sky-line regions. The line flux limit in these regions is similar, at \mbox{$\leq1\times10^{-16}$\,ergs\,s$^{-1}$\,cm$^{-2}$}. Since the continuum is so poorly detected, there are no useful constraints on the line equivalent widths. 

The stellar size of $C$ is constrained at high SNR in $K_{\rm s}$ band, with an unresolved Gaussian FWHM fit of $(0.59\times0.53)$\,arcsec$^2$ , implying a $<2$\,kpc half-light radial size. For a massive galaxy at early epochs this is not particularly unusual \citep{Damjanov}, 
however, as an early  BCG galaxy this is tiny. By $z\approx1$, half-light radii for BCGs in massive clusters range from 14 to 53\,kpc, 
with an average determined from stacking of $32.1\pm2.5$\,kpc for $z\approx1$ \citep{Stott11}. (We discuss SPT2349-56 relative to $z=1$ clusters below.)
This is only 30 per cent smaller than the size in low-redshift comparison samples with $43.2\pm1.0$\,kpc. 
Thus the SPT2349-56 progenitor BCG must grow in size dramatically over the subsequent $\approx2$\,Gyrs, for example through dry mergers \citep{Cooke19}.  

\begin{figure}\centering
	\includegraphics[width=9.2cm]{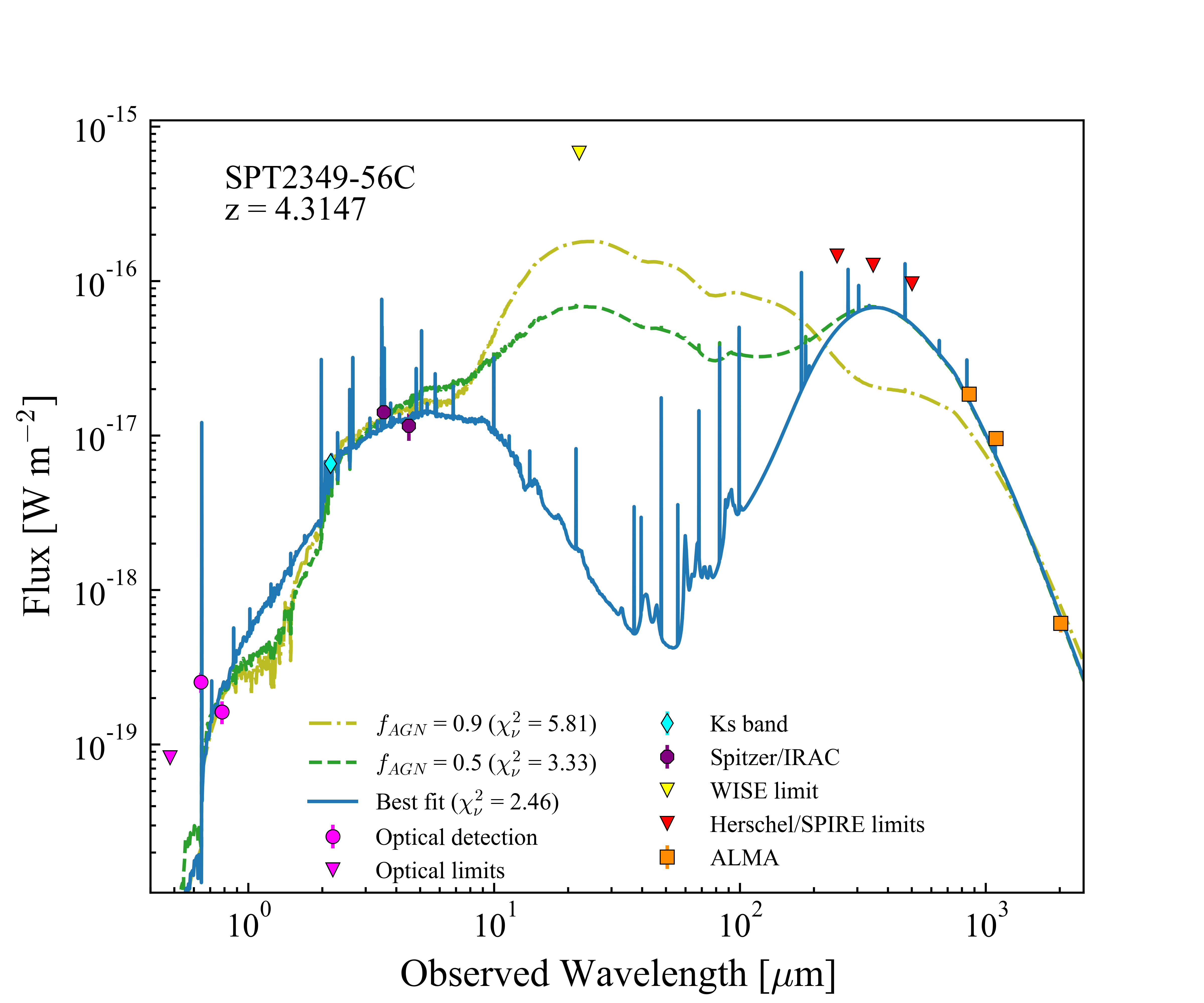}
	\caption[\textsc{cigale} fit of source $C$]{\textsc{cigale} fit to source $C$, the brightest at rest-optical wavelengths and the only ALMA source with high significance detection in the $K_{\rm s}$ band. The best fit includes a negligible AGN fraction, with an excess in $3.6\,\mu$m potentially driven by strong H$\alpha$ emission.
	Higher AGN fraction fits are shown to demonstrate that the available mid-IR data cannot constrain the AGN contribution in this source.}
	\label{fig:2349CIGALE}
\end{figure}

In addition to the stellar mass fitting for source $C$, we attempted full SED fitting of UV-through-mm wavelengths. Dust emission is modelled using the updated empirical templates from \cite{Draine14} (originally from \citealt{Draine07}). Limits from {\it Herschel}-SPIRE are adopted at $250{-}500\,\mu$m \citep{Miller}, effectively restricting the range of templates possible.
The best fit has the following parameters: a stellar mass of 
$(3.2^{+2.3}_{-1.4})\times10^{11}$\,\msun;
a SFH age of the main population of $(1.1\pm0.26)$\,Gyrs; 
and an AGN fraction of 0.0 with an upper limit of 0.4. The best-fitting SED is shown in Fig.\,\ref{fig:2349CIGALE}.

We attempted to constrain the AGN contribution using templates from \cite{Fritz06}. However, the only mid-IR constraint is from WISE at $22\,\mu$m, with supplemental SPIRE limits, which were not sufficiently deep to quantify the fractional AGN contribution. During the fitting, we allowed the AGN fraction to vary from 0.0 to 0.9 in increments of 0.1 and let the AGN be either type-1 or type-2. A fractional contribution of close to 0.0 is preferred based on the probability distribution function and the reduced $\chi^2$, but as evident from Fig.\,\ref{fig:2349CIGALE}, higher AGN fractions are also possible and more data are required to further constrain this.

In contrast, source $C$ does have some properties suggestive of an AGN at other wavelengths. It has a much narrower \CII\ line profile than all other similarly bright sources in the proto-cluster, and accordingly one of the lowest \CII/$L_{\mathrm{FIR}}$ ratios found in the structure \citep{Miller,Hill20}, typical of an AGN \citep{Stacey10}. It has a large CO(16--15) luminosity with a CO excitation more consistent with 
AGN-dominated galaxies (Canning in prep.). 
However, our SED fitting appears consistent with a large stellar mass, and does not obviously require an AGN component from a hot dust torus \citep[e.g.][]{Hainline11}. Further, the 3.6-$\mu$m excess is well modeled with a strong H$\alpha$ line component, which JWST observations will be able to confirm.

\section{Discussion}

\begin{figure}\centering
	\includegraphics[width=8.2cm]{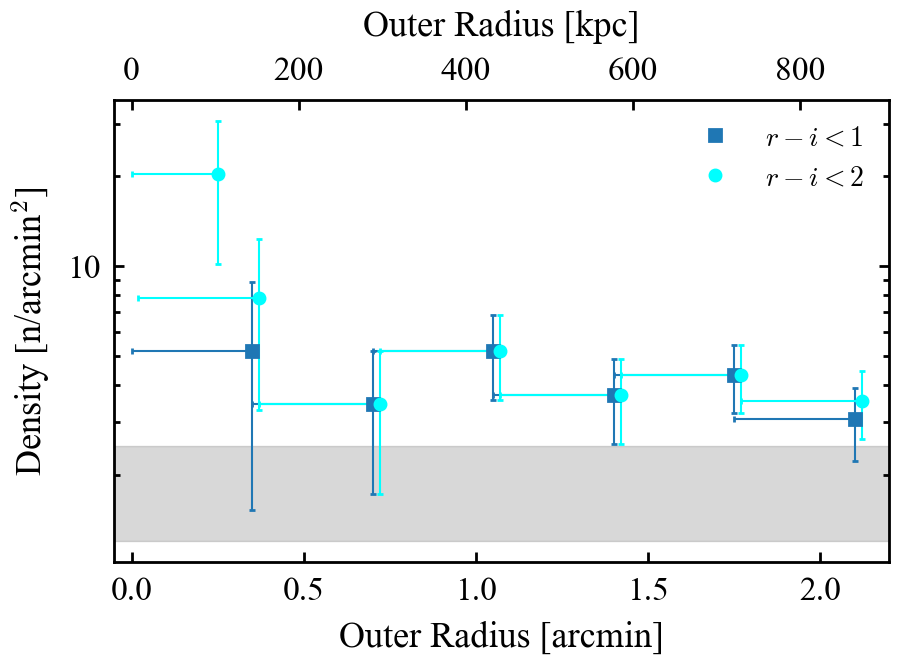}
	\caption[]{Number density of $z\simeq4$ LBGs with $i\lesssim26.2$ (${>}5\sigma$ detections) 
	within a 2.1\,arcmin radius of the ALMA centroid, determined for radii of 0.35\,arcmin increments. Each annulus has an inner radius 0.35\,arcmin smaller than the outer radius. The number density for $z\simeq4$ LBGs with the \cite{Toshikawa18} colour criteria are shown as blue squares, while the cyan circles (offset for ease of viewing) show the number density when the upper limit of the $r-i$ colour is relaxed to 2. The cyan circle at 0.25 arcmin radius shows the number density of the 4 LBGs within the central 30\,arcsec-diameter core of SPT2349-56 discussed in the text. The range in field number density extrapolated from \cite{Toshikawa18} and \cite{Steidel99} is identified by the grey region.
	}
	\label{fig:2349lbghist}
\end{figure}

\subsection{Implications for optical proto-cluster identification} 
We have demonstrated that the SPT2349-56 proto-cluster core is difficult to study in either its rest-UV or rest-optical properties, due to the extreme faintness of most of its members.
A parallel study with the VLT/MUSE integral field spectrograph has also demonstrated that these ALMA sources are not detected in Ly$\alpha$, although there appears to be an overdensity of Ly$\alpha$ emitters in the surrounding $\approx1$\,arcmin field (Apostolovski in prep.).

It is of interest to determine whether this proto-cluster would be detected in a large optical survey using the same $g$-band dropout selection criteria we used to identify LBGs within the SPT2349-56 core. Our result in SPT2349-56 can be compared to the recent search for proto-clusters across a 121\,deg$^2$ optical survey using Subaru's Hyper Suprime-Cam \citep{Toshikawa18}, which reached somewhat shallower depths in the $g, r$ and $i$ bands compared to  our Gemini-S data. \cite{Toshikawa18} selected galaxies around $z\approx4$ using the $g$-band dropout technique that we have adopted here, and searched for proto-cluster candidates by computing the number of these LBGs within 1.8\,arcmin-radius apertures (about 0.75 proper Mpc). They found that the mean number of LBGs to $i_{\rm AB}<25$ within such an aperture is 6.4, with a standard deviation of 3.2. By adopting a $4\sigma$ overdensity threshold, a large number of proto-cluster candidates were found. 
Adopting the same metric to \cite{Toshikawa18}, we find 16 LBGs to $i_{\rm AB}<25$ in a 1.8\,arcmin-radius aperture for SPT2349-56, which would be a marginal $3\sigma$ overdensity detection and would therefore not meet the threshold set by \cite{Toshikawa18}. 

By going to deeper limits, we can improve the statistics. Extrapolating the mean LBG density around $z=4$ of \cite{Toshikawa18} (using the luminosity function of \citealt{Steidel99}) to $i_{\rm AB}\lesssim26.2$ (corresponding to ${>}5\sigma$ detections in our $i$-band image) results in a field density of about 2.5\,arcmin$^{-2}$. \cite{Steidel99} find a number density of just 1.2\,arcmin$^{-2}$ in their survey area of 0.23\,deg$^2$ (with slightly different filters). Given the significantly larger area of \cite{Toshikawa18} we estimate the following overdensities in comparison to \cite{Toshikawa18}.  
We have found four LBGs within a 30\,arcsec-diameter region surrounding the centre-of-mass of SPT2349-56 and encircling the brightest 14 ALMA sources in its core. This corresponds to an overdensity of about 8 times the background (at most 
17 times), with a large sensitivity to the enclosed region chosen (taking the smallest possible circle that encloses the four LBGs 
results in a overdensity of 15 times the background, at most 32 times). 
The fact that even one LBG at this depth is only expected $\approx$10 per cent of the time within a circular area of about 0.2\,arcmin$^2$ reinforces the likelihood that all four of these LBGs lie in the SPT2349-56 structure. 
The LBG density appears to fall off to around $3{-}4$\,arcmin$^{-2}$ outside this core region, remaining  constant within uncertainties 
out to $\simeq2$\,arcmin radial distances (see Fig. \ref{fig:2349lbghist}). 
Thus, while the SPT2349-56 core contains a large overdensity of LBGs, 
only a modest overdensity of $2{-}3$ times the background remains outside the central 30\,arcsec-diameter region. In the outer region, the number densities are also consistent within errors when considering the LBGs selected using either the $r-i<1$ or $r-i<2$ colour windows. However, a more significant difference (a factor of 2$\times$) in the overdensity is observed in the core region when considering these relaxed LBG selections (see Sect.~3.1). Thus, we  find evidence that fainter and redder LBGs are signficantly overdense in the core, suggesting a higher dust obscuration in galaxies near the centre-of-mass of the proto-cluster.

As noted in \cite{Steidel99}, the locus of unevolved early-type galaxies at $z=0.5-1$ comes very close to the selection window for $z\approx4$ galaxies, and a combination of photometric errors and intrinsic variations in galaxy SEDs may scatter early-type galaxies at these low redshifts into our LBG selection window. Interlopers are identified by \cite{Steidel99} as those objects in the selection window with typically redder $\mathscr{R}-I$ colours. A red $\mathscr{R}-I\leq1.2$ criterion is expected to result in a 34 per cent contamination rate as opposed to 20 per cent for the more stringent $\mathscr{R}-I\leq0.6$ selection window. While the \cite{Steidel99} filter bands differ from our own, the contamination rate is likely similarly affected.
In the core region of SPT2349-56, the relaxed colour criteria have undeniably added true $z\approx4$ galaxies, despite the possibility of increased interlopers. In the outer region the number of additional LBG candidates found with redder colours is statistically insignificant, with only two identified within a 2\,arcmin radius. 

Our conclusions about the detectability of SPT2349-56 in optical surveys would apply equally to surveys from the Rubin Observatory, which will reach similar depths as Subaru's Hyper Suprime-Cam over large areas \citep{Robertson19}. While such large-area optical photometric surveys will identify larger numbers of proto-cluster candidates, they do not improve the sensitivity to structures like SPT2349-56. 
This in itself is not surprising, given the broad redshift range $z=3.3{-}4.2$ of the $g$-band dropout selection, which makes it a relatively blunt tool to identify overdensities without spectroscopic follow-up \citep[e.g.][]{Steidel96} -- even relatively strong spikes in redshift distributions characteristic of large overdensities \citep[e.g.][]{Steidel00} do not manifest themselves as strong overdensities in the photometric LBG selection.
%
In fact, the large overdensity of LBGs at $z=3.09$ in the SSA22 field \citep{Steidel98} and at $z=2.3$ near QSO HS1700+643 \citep{Steidel05} only appear as significant overdensities through spectroscopic redshift analysis. 
In simulations of proto-clusters at $z=3$, \cite{Chiang13} also noted that for photometric colour-selection surveys, where the redshift uncertainty is sufficiently large ($\Delta z\gtrsim0.1$), the galaxy overdensity is essentially indistinguishable from the field density, except for the most overdense systems. Optical surveys therefore result in only a partial sample of proto-clusters. The fact that SPT2349-56 does not meet the \cite{Toshikawa18} $4\sigma$ overdensity threshold, does not preclude it from being a true proto-cluster.
%
This highlights that a search for proto-cluster candidates is incomplete if  surveys of only spatial overdensities of photometrically selected LBGs 
are considered. Due to the challenges associated with determining spectroscopic redshifts at high-$z$, detecting overdensities of spectroscopically confirmed LBGs only becomes more difficult at $z=4.3$.

On the other hand, simple arguments demonstrate that in large optical surveys for $z\simeq4$ dropout-galaxies \citep[e.g.][]{Toshikawa18}, the four central LBGs in SPT2349-56 are not unusual.
In the absence of clustering, even a 1\,deg$^2$ survey has a ${>}\,50$ per cent chance to find four $z\simeq4$ LBGs to our depth in a 30\,arcsec-diameter circle (0.2\,arcmin$^2$).
Thus we conclude that optical surveys are typically quite blind to structures like SPT2349-56. Nevertheless, SPT2349-56 clearly represents an early phase of one of the most massive structures in the Universe, a truly dust obscured phase of a massive cluster core under formation.

In their analysis of cosmological simulations, \cite{Muldrew15} find proto-clusters to be very extended objects -- with 90 per cent of their total mass spanning an approximately 60\,arcmin-diameter region by $z=2$, the angular size remaining largely constant from $z=1{-}5$. Additionally, 90 per cent of the stellar mass resides in an area with a diameter ranging from $14{-}24$\,arcmin, depending on the halo mass of the proto-cluster. They suggest that the vast majority of proto-cluster surveys, whose fields of view are typically significantly smaller, therefore miss a large percentage of the proto-cluster mass. Specifically, they hypothesize that only the densest proto-cluster cores are being observed. Furthermore, \cite{Muldrew15} find that proto-clusters are generally comprised of multiple haloes, only a small fraction of the stellar mass (about 20 per cent) residing in the most massive halo (or the proto-cluster core) observed by most studies. SPT2349-56, whose incredibly active core is encompassed in a mere 
0.5\,arcmin-diameter region, is therefore significantly more compact than the `typical' simulated proto-cluster core.

While optically identifiable sources are found in the proto-cluster, distinct from the ALMA-identified sources, 
our analysis reinforces the notion that this incredibly massive and active structure is most easily identifiable in mm-wave surveys like that of SPT. 
%
AGN are important in the evolution of clusters and BCGs, providing feedback and regulation to the star formation evolution. However, no bright quasars/AGN candidates are found in the region from optical colour selection. Moreover, our imaging of these 14 SMGs do not reveal any obvious signs of AGN contribution, even in the bright SMG $C$.
Finally, the foreground galaxy at $z=2.54$ along the line-of-sight to ALMA source $A$ illustrates the importance of multi-wavelength analysis and spectroscopic follow-up in such deep observations of crowded structures.

\begin{figure*}\centering
	\includegraphics[width=0.9\textwidth]{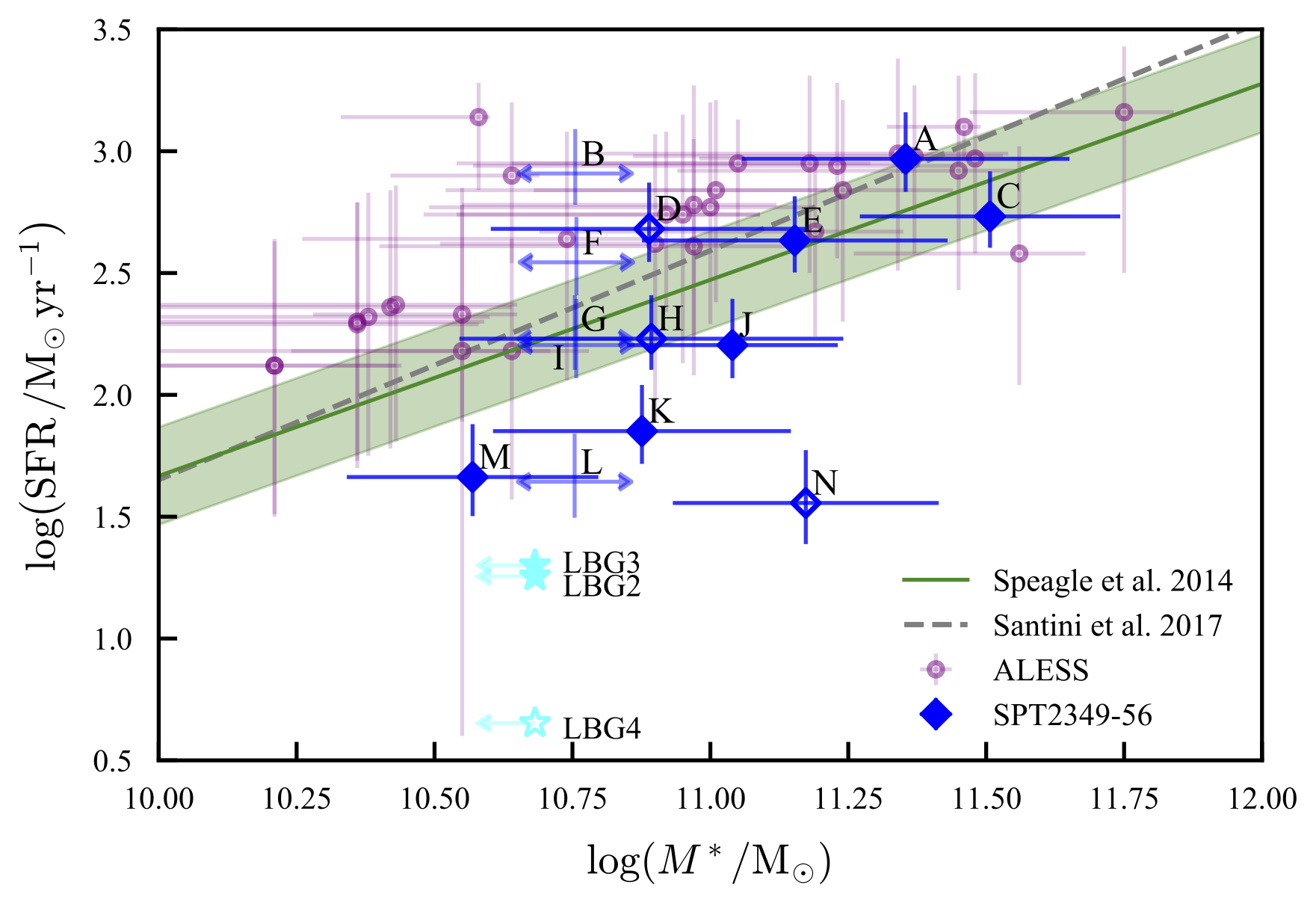}
	\caption[Main sequence star-forming galaxies]{Assessing the SPT2349-56 sources relative to the main sequence (MS) of star-forming galaxies. 
	SPT2349-56 SMGs are shown with \mstar\ values derived from \textsc{cigale} best fits (blue diamonds -- open symbols have additional constraints on the fits as described in text).
	Constraints on \mstar\ for those lacking optical/nearIR detections (blue double-sided limits) are shown at the scaled IRAC limits as described in the text.
	However the non-detections may be driven by extreme extinction since their dynamical and gas masses 
	are generally comparable to the SMGs detected by IRAC, and thus we show right-sided limits as well to reflect this.
	LBGs are shown as 3$\sigma$ IRAC limits scaled as described in text. 
	The SFRs for the SMGs are scaled from integrated $L_\text{FIR}$ estimates \citep{Hill20}. SFRs for the LBGs are estimated as described in the notes of Table~2. 
	The $z>3.5$ SMGs \citep{daCunha15} from the ALESS survey \citep{Simpson15} are shown as purple circles.
	The 
	MS at $z=4.3$ 
	determined by \cite{Speagle14} is shown in green, where the scatter (shaded green) is $\pm\,0.2$ dex.
	The MS power law in the $4\leq z<5$ range from \cite{Santini} is also shown (grey dashed line).
}
	\label{fig:2349ms}
\end{figure*}

\subsection{Main sequence of star formation}
Having estimated the stellar masses, we can assess the relation between the SFR and stellar mass for the SPT2349-56 SMGs, adopting SFRs based on ALMA photometry from \cite{Hill20}, and compare them to the coeval field population. For a given redshift, the majority of star-forming field galaxies are observed to exhibit a correlation between these two properties \citep[e.g.][]{Noeske,Speagle14,Santini}. \cite{Santini} suggest that the tightness of the correlation, defined as the main sequence (MS), is due to similarities in the gas accretion histories. For galaxies along the MS it is expected that the dominant mechanism for growth is a smooth accretion of gas from the intra-galactic medium over long timescales. 
Bright SMGs have been proposed to lie significantly above these correlations \citep[e.g.][]{Hainline11, Michalowski12}, possibly due to major mergers triggering intense SFRs \citep[e.g.][]{Engel10}. 
The empirically defined MS is typically parameterized as a power law,
\begin{align}
	\log{(\mathrm{SFR}/\mathrm{M}_{\odot}\,\mathrm{yr}^{-1})}=\alpha\log{(M^*/\mathrm{M}_{\odot})}+\gamma,
\end{align}
where $\alpha$ is the slope of the $\log{\mathrm{SFR}}-\log{M^*}$ relation, and $\gamma$ is the MS normalization. While the normalization is clearly observed to increase with redshift, the slope is still debated. Generally it is believed that the slope is predominantly unevolving, and approximately linear in power-law near unity. For the $4\leq z<5$ range, \cite{Santini} find $\alpha=0.94\pm0.06$ and $\gamma=1.37\pm0.05$. 
\cite{Speagle14}, through a literature review in which 25 studies were considered, determined a 
MS best fit with $\alpha=0.80\pm0.02$ and $\gamma=6.4\pm0.2$ when the age of the Universe is set to 1.35\,Gyrs ($z=4.3$), with a scatter of $\pm\,0.2$ dex. 
While \cite{Speagle14} did not include data from the first and last 2 Gyrs of the Universe in their 
fitting, they find their MS relation to provide a reasonable fit to the data even out to $z\sim5$ and that the parameters of the MS are only marginally affected when including high-$z$ data.

Figure\,\ref{fig:2349ms} shows the relation between SFR and \mstar. 
The 
MS from \cite{Speagle14} is shown for a redshift of 4.3, as well as the \cite{Santini} 
$4\leq z<5$ MS. Five of the eight IRAC-detected sources ($A,C,D,E,$ and $J$) as well as the $K_{\rm s}$-detected source $H$ appear to lie within the scatter of the MS of \cite{Speagle14}.
This assessment is unchanged  if the \cite{Santini} MS relation is instead adopted.
Several of the five IRAC-undetected sources (especially $B$ and $F$) could lie above the MS if their inferred stellar masses are truly as low as the flux limits suggest, but extreme extinction may be driving the faintness rather than a low \mstar. 
Indeed, their dynamical and gas masses (from $^{12}$CO line widths and $^{12}$CO luminosities, respectively) are generally as large as those sources detected by IRAC, which would imply their stellar masses are similarly large, unless they have very high gas fractions relative to stars.  Our gas fraction analysis above does suggest that the stellar masses may well be as large or even larger than the location of our IRAC limits would imply (see Table~\ref{table:dataSMGprop}). We thus show these five SMGs as double-sided limits to reflect this possibility.

We further compare the SPT2349-56 SMGs in Fig.~\ref{fig:2349ms} to a sub-sample of the \cite{daCunha15} isolated SMGs from the blank field ALESS survey \citep{Simpson15}, where we have restricted the redshift range to $z>3.5$ (a mix of spectroscopic and photometric redshifts). The final maps of the ALESS survey have a median RMS of $\sigma_{870\,\mu\rm{m}}=0.21$ mJy beam$^{-1}$ \citep{Simpson15}. The ALMA maps at $850\,\mu$m presented by \cite{Hill20} have a depth of 0.03 mJy beam$^{-1}$, a factor of about $7\times$ deeper. 
The \mstar\ values in the 10 SPT2349-56 sources with higher SFRs (SFR $>100$\,\msunrate, a limitation set taking the continuum image depths into account to facilitate a more appropriate comparison with ALESS) 
likely have a similar median stellar mass to those of ALESS, $0.9\times10^{11}$\,\msun. Because four of these 10 SPT2349-56 SMGs only have limits, we cannot probe this comparison further.

Interestingly, the four SPT2349-56 SMGs with SFRs below the detection limit of ALESS have similar \mstar\ values to the much higher SFR ALESS galaxies, suggesting they either have atypically high stellar masses or have SFRs well below the MS given their stellar mass.  
%
At least three ($K, M$ and $N$) and plausibly all four (given that $L$ is a limit) lie significantly below the MS.
While quenched galaxies at $z>4$ are exceedingly rare \citep[e.g.][]{Speagle14}, 
environmental factors in the dense proto-cluster core may have accelerated the quenching of these galaxies. `Preventative' and `ejective' feedback mechanisms are methods in which the gas reservoir of a galaxy is either heated or expelled, thereby inhibiting the conditions required for further stellar activity, effectively quenching galaxy growth. While the location of these SPT2349-56 SMGs below the MS is suggestive of quenching, we can only speculate on the mechanism. Further, it is also not clear whether various quenching mechanisms operate differently on specific mass ranges. 

In our comparison with the $z=4{-}5$ main sequence of star formation, we find evidence for SMGs at three potentially different stages of evolution. 
The fact that many of these SMGs are all apparently well situated along the MS reveals that even in the most extreme environment ever found at $z>4$, 
star-forming galaxies are not clearly offset from the scaling relations of coeval field galaxies. Indeed, \cite{Speagle14} find the mode of star formation at a given mass to be independent of the density of the environment.
%
SMGs lying above the MS may be gas-rich galaxies at early evolutionary stages, which are driven to high SFRs through the dense merger environment of the core.
%
%
However, in the simulation conducted by \cite{Rennehan} of the evolution of the 14 SMGs in the proto-cluster core, 
the intense star formation and resulting stellar feedback together dramatically reduce the gas in the merger system resulting in rapid decline in star formation.
The SMGs below the MS may represent quenched galaxies whose star formation has been truncated through feedback in this environment. 
%
The LBGs identified in SPT2349-56, distinct from SMG identifications, are not obviously very massive galaxies. They may require much deeper near-IR observations to better constrain their stellar masses.
The considerable diversity of these sources in their evolutionary phases is no doubt a result of the extreme merger environment of these 14 SMGs, which are contained within a 130\,kpc-region, no larger than the dark halo of the Milky Way! 

\subsection{\bf Why are there so many active SMGs in the core?
}

It is clearly a 
rare phenomenon to observe 14 star-forming galaxies in such a compact 
region. 
Our SED fitting has allowed for some constraints on the stellar ages and star formation histories of the SMGs (Table~3), which in principle could help to elucidate why there are so many active, bright SMGs in this proto-cluster. 
Unfortunately, our photometric sampling is too sparse and our measurements often too low a SNR for precise SED fitting that would allow us to discuss in detail the star formation histories. We defer this analysis to Hill (in prep.), as well as a more detailed consideration of the duty cycles and gas masses. 

We can however speculate on the reasons for the number of bright, synchronous SMGs in this proto-cluster core, bringing our findings of additional optically detected proto-cluster members to the picture.
SMGs are a rare population of galaxies that are detected because of their high sub-mm flux density, related to their short-lived, elevated SFRs. An inherent stochasticity then likely  exists in selecting massive galaxy haloes hosting SMGs \citep{Miller15}, in that a given halo may or may not be seen as active depending on the duty cycle of these star forming bursts. \cite{Miller15} conduct counts-in-cells analysis of eight mock SMG catalogues (each with about a 2\,deg$^2$ field of view from $z=0.5-8$), choosing cell sizes of 10\,arcmin\,$\times$\,10\,arcmin with depths of ${\rm d}z=0.05$, ensuring that no SMG associations are split by cell boundaries and that no associations are counted twice. The simulations 
indeed show that while the largest associations of SMGs (five or more) do trace the massive dark matter haloes likely to evolve into present day rich galaxy clusters, these associations are 
uncommon (only about 1 per cent of SMGs exist in such large associations). Furthermore, many of the most massive overdensities do not contain any SMGs,  in the $z=2-4$ range. 

The median pairwise separation of SMGs in the largest simulated associations is 4.3\,Mpc; consequently \cite{Miller15} 
concluded that no environmentally driven star formation could be the cause of these SMG associations. 
However, the 14 SMGs of SPT2349-56 are all located within a region approximately $35\times$ smaller, and we have here further uncovered another three relatively massive galaxies in the same volume with lower SFRs. \cite{Hill20} have also recently found several \CII\ emitters in the same volume with low SFRs. The extremely dense environment of SPT2349-56 results in many of the SMGs being close enough to interact on roughly dynamical timescales -- through mergers, shocks, and energetic outflows from star formation and AGN, as shown in the simulations of \cite{Rennehan}. This, possibly along with dense filamentary inflows of gas, may lead to the global ignition of star formation witnessed here occurring within a ${\sim}\,130$\,kpc-region.

SPT2349-56 is the brightest example in the SPT-PC survey. With just nine proto-cluster candidates in the $2,500$\,deg$^2$ SPT field (Wang et al. submitted), they are exceptionally rare, with number densities barely reaching $n\approx0.004\,\rm{deg}^{-2}$. While it is unclear whether the star formation recipes in \cite{Miller15} could ever result in a system like SPT2349-56, it is conceivable that the volume probed in the 15.7\,deg$^2$ field of view of \cite{Miller15} is simply too small to detect any SMG associations comparable to that seen in SPT2349-56.

\subsection{Stellar growth of the BCG} 
Source $C$ stands out in Fig.\,\ref{fig:2349ms}, with an inferred \mstar\ as massive as any SMGs seen in the literature at any redshift. 
SED fitting is consistent with $M^*=(2{-}6)\times10^{11}$\,\msun\ (from the width of the \textsc{cigale} probability distribution) 
and the gas-rich merging complex seen in sources $B,C,$ and $G$ may represent the accumulated stellar population of a forming BCG in the core of a massive galaxy cluster. 
The cumulative stellar mass for all the SMGs in the SPT2349-56 core is at least $(12.2\pm2.8)\times10^{11}$\,\msun\ (considering only nine of the 14 SMGs are included), which is comparable to the gas mass of the 14 SMGs derived in \cite{Miller}, $6.7\times10^{11}$ ($\alpha_{\rm CO}$/0.8)\,\msun. 
This gas mass is normalized to the conservative CO-luminosity-to-gas-mass conversion factor
typically adopted for very luminous galaxies \citep[e.g.][]{Tacconi10}. A more likely estimate would scale $\alpha_{\rm CO}$ continuously towards $\alpha_{\rm CO}=4$ for the lowest SFR (and mass) galaxies \citep{Narayanan10}, yielding $\sim10^{12}$\,\msun.

Since \cite{Rennehan} have estimated that all 14 galaxies will completely merge in around 500\,Myrs and that the stellar mass will increase in this time due to the partial (50 per cent) consumption of this gas, we can infer that by a redshift of only 3.3 the stellar mass of this assembling BCG could be in excess of $1.5\times10^{12}$\,\msun. 
%
\cite{Collins09} and \cite{Tonini12} find that the vast majority of the stellar mass in BCGs is already in place at very early times; 50 per cent by $z=5$ and 80 per cent by $z=3$. \cite{Tonini12} consequently also find that the time scale of the mass assembly and the age of the bulk of the stellar component are comparable, roughly 2-3 Gyrs, suggestive of monolithic collapse.

\begin{figure}\centering
	\includegraphics[width=0.48\textwidth]{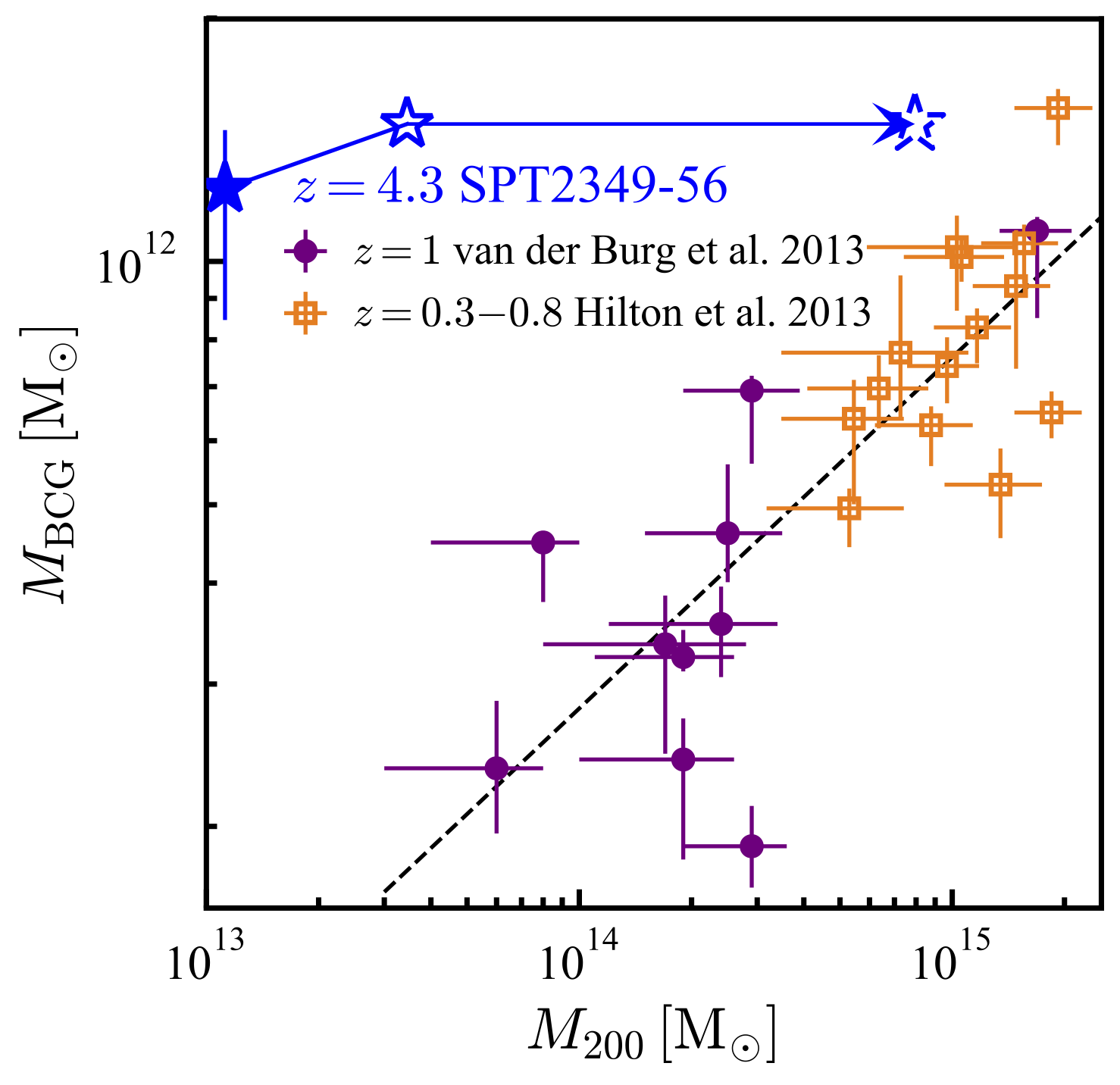}
	\caption[summary mstar]{The assembling BCG stellar mass of SPT2349-56 compared to a sample of $z\simeq1$ galaxy cluster data from \cite{vanderBurg13} (purple circles), and $z=0.3{-}0.8$ clusters from \cite{Hilton13} (orange squares). We show the summed \mstar\ of the nine detected SPT2349-56 SMGs (blue filled star) and the predicted increase of \mstar\ in 500\,Myrs to $z=3.3$ (open blue star) from \cite{Rennehan}.  A growth in $M_{200}$ of $3\times$ is expected from $z=4.3$ to $z=3.3$, and  $70\times$ from $z=4.3$ to $z=1$ \citep{Chiang13}, providing an evolutionary track to $z=1$ assuming no additional growth in \mstar\ of the SPT2349-56 BCG (dashed blue star). The dashed black line is a fit to the joint cluster datasets.
	}
	\label{fig:bcg}
\end{figure}

In Fig.\,\ref{fig:bcg}, we show the assembling BCG stellar mass of SPT2349-56 compared to a sample of $z\simeq1$ galaxy cluster data from \cite{vanderBurg13}, and $z=0.3{-}0.8$ clusters from \cite{Hilton13}. 
The summed \mstar\ of the nine detected SPT2349-56 SMGs is already comparable to that of BCGs of the most massive $z=1$ clusters (${>}\,10^{15}$\,\msun\ halo mass), and strongly suggests an inside out collapse of a very massive structure \citep{vanderBurg15}, with such an accelerated growth of the core mass relative to $z=1$ clusters.  
We show the summed \mstar\ of the nine detected SPT2349-56 SMGs  and the predicted increase of \mstar\ in 500\,Myrs to $z=3.3$  from \cite{Rennehan}.  A growth in $M_{200}$ of $3\times$ is expected from $z=4.3$ to $z=3.3$ (the predicted 500\,Myrs depletion time of $M_{\rm gas}$ mentioned above) and  $70\times$ from $z=4.3$ to $z=1$ \citep{Chiang13}. 
Starting from the virial mass of the SPT2349-56 core \citep{Miller}, this provides a rough evolutionary track for SPT2349-56 leading to $z=1$, assuming no additional growth in \mstar.
In hierarchical $\Lambda$CDM modelling, \cite{Ragone18} find that on average the stellar mass of the BCG and the extended stellar halo grows from $z=2$ to $z=1$ by a factor of about 1.6 with an additional factor of 1.5 from $z=1$ to $z=0$.
Even without accounting for the stochastic growth associated with merging satellites as in the \cite{Ragone18} simulations, SPT2349-56 ends up at $8\times10^{14}$\,\msun\ in $M_{200}$ by $z=1$, with a significant excess in BCG mass, relative to the comparison cluster samples (although not a vastly unusual outlier given the dispersion). 
The fact that SPT2349-56 already appears to have an excess of stellar mass at $z=4.3$, and that further rapid BCG growth is necessitated by the merger and enormous gas mass present in these galaxies,  suggest it may well be an outlier in the BCG / $M_{200}$ relation by $z=1$.

\begin{figure}\centering
	\includegraphics[width=0.48\textwidth]{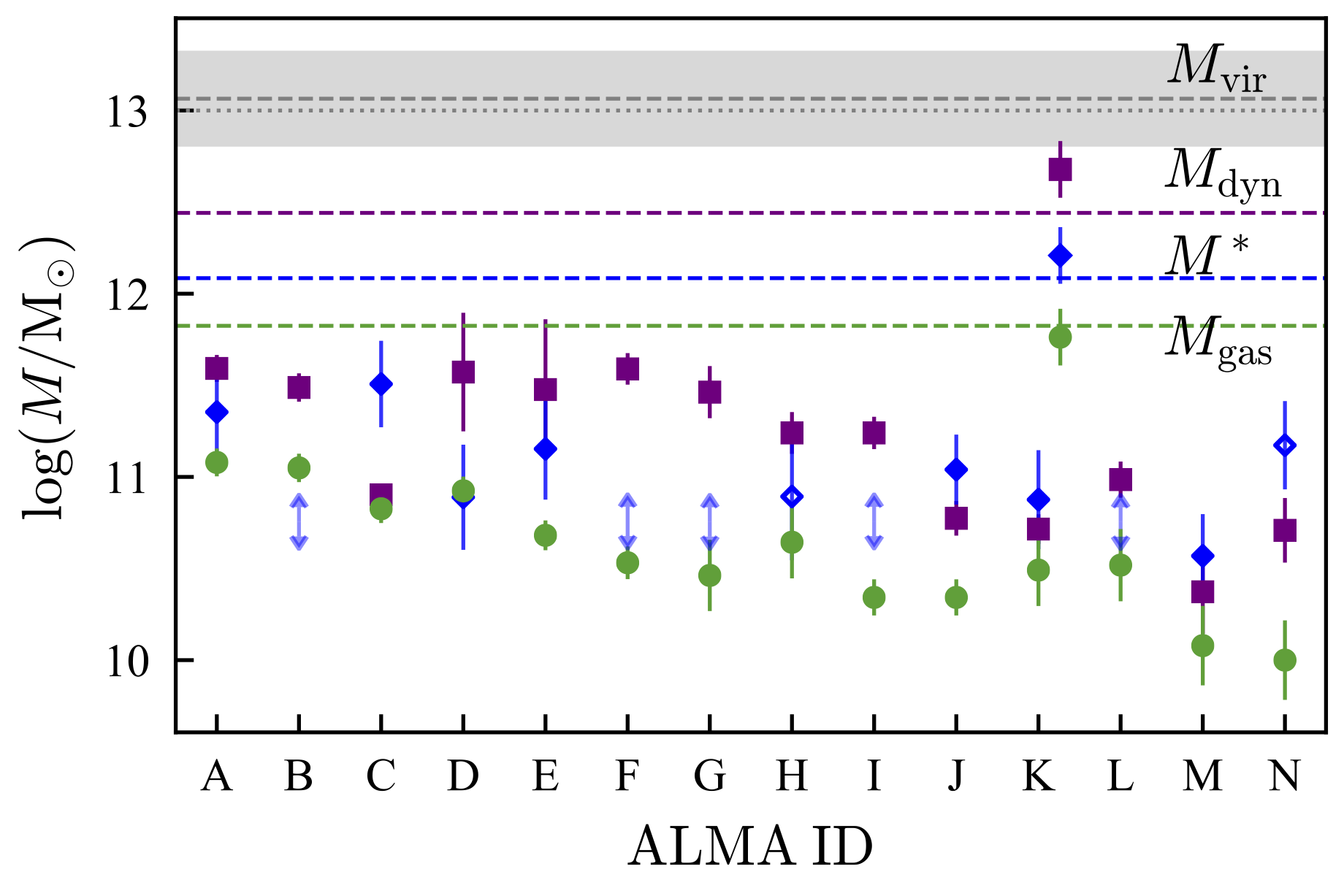}
	\caption[Masses]{Mass budget for the individual sources in SPT2349-56. Stellar masses (blue diamonds) are determined from SED fitting as discussed in Sect.~3.2. The cumulative stellar mass (blue dashed line) only includes the nine sources with constrained SED fits. We adopt the gas masses (green circles) from \cite{Miller}. Dynamical masses (purple squares) are estimated within a 2\,kpc-radius given the \CII\ line widths of \cite{Miller}, using the equation presented in Sect.~3.1.1. Cumulative gas and dynamical masses are shown with  green and purple dashed lines, respectively. The grey dashed line \citep{Miller} and grey dotted line \citep{Hill20} are the virial mass estimates for the proto-cluster.
    }
	\label{fig:masses}
\end{figure}

\subsection{Baryon budget of SPT2349-56}
Next, we turn to an assessment of the baryon budget of SPT2349-56 at $z=4.3$ and explore whether some of the SPT2349-56 baryons may already be in the form of a diffuse, hot gas filling the space between the galaxies -- the intra-cluster medium (ICM) that is characteristic of massive virialized galaxy clusters at $z<1.5$.  
The virial mass of these 14 SMGs is estimated as $(1.16\pm0.70)\times10^{13}$\,\msun\ in \cite{Miller}.
\cite{Hill20} find a slightly lower $M_{\rm vir}\simeq1\times10^{13}$\,\msun\ from a larger number of galaxies within a somewhat larger radius, but still well consistent within the uncertainties. 
In Fig. \ref{fig:masses}, the stellar, gas, and dynamical mass estimates for the 14 SPT2349-56 sources  are shown individually, as well as their cumulative masses in comparison to the virial mass of the proto-cluster. 
%
We note that several SMGs (sources $C,J,K,M$ and $N$) have  $M^*>M_{\rm dyn}$. 
This is not surprising considering the large uncertainties associated with the stellar mass estimates, up to about 20 per cent. Furthermore, in the absence of radial size estimates for each SMG, we assumed an average 2-kpc radius in the dynamical mass calculations. 
For sources $K,M,$ and $N$, which lie significantly below the main sequence (source $J$ only marginally), the dynamical mass is comparable to the stellar mass within error while the gas masses are depleted, consistent with our suggestion that these galaxies have been quenched. 

Under our assumptions for estimating these quantities, the ALMA galaxies on the whole appear to be mostly dark matter dominated ($M_{\rm dyn}>M_{\rm baryon}$). Combining the stellar and gas masses of these core SMGs, we have a total cold baryon mass of at least 
$1.9\times10^{12}$\,\msun, and a total dynamical mass of 
$2.8\times10^{12}$\,\msun, suggesting on average the galaxies have a dark matter fraction of about 30 per cent within this assumed 2\,kpc-radius.

We can divide the total baryon mass by the virial mass estimated by \cite{Hill20}, which yields a cluster baryon fraction of $f_{\rm b}=0.19$.
This is comparable to  the universal baryon fraction estimate based on Planck determinations of $\Omega_{\rm b}/\Omega_{\rm m}=0.156\pm0.003$ (Planck Collaboration XIII, 2016). Notice that stars by themselves give baryon fractions of $f_{\rm b}>0.12$.
For the five IRAC-undetected SMGs, the dynamical masses (from $^{12}$CO line widths) are generally as large as those for the SMGs detected by IRAC. This may imply their stellar masses are similarly large, but the SEDs have large dust extinctions, which would suggest the universal baryon fraction is exceeded just by the stellar masses of the SPT2349-56 SMGs. Therefore, unless the gas and stellar masses are dramatically overestimated, even with uncertainties, there is little room for significant amounts of hot, diffuse ICM gas. \cite{Sokolowska} show that this is consistent with recent high-resolution galaxy simulations, which suggest that at the earliest stages the growth of the diffuse gas component is primarily driven by galactic outflows powered by stellar (and AGN) feedback, as well as heating due to major mergers. 

\section{Conclusions}

We have studied SPT2349-56 at optical and near-IR observed-frame wavelengths with imaging from Gemini-S and {\it Spitzer}-IRAC. Arguably the most important conclusion from this work is that despite the incredible total SFR and density of SPT2349-56, it would be exceedingly difficult to identify in large surveys through optical overdensity selection techniques, and the structure is faint or undetected even at near-IR through IRAC wavelengths. This emphasizes the importance of searching for early formative structures at millimetre wavelengths. 
Also of importance for cluster formation, we find the likely BCG associated with ALMA source $C$. This is significant, as it coincides with where most of the \CII\ and CO(4-3) emitters are spatially located, and significant stellar growth has already occurred near the centre-of-mass of the cluster core.
%
Our conclusions are as follows.
\begin{itemize}
\item We detect four plausible counterparts to SPT2349-56 SMGs in the $g,r,$ and $i$ bands, although one is revealed to be an interloper $z=2.54$ galaxy along the line-of-sight to SMG $A$. We estimate a possible 
resulting gravitational lensing magnification of 
ALMA source $A$ to be $<1.2$.

\item Using the \cite{Toshikawa18} 
$z\simeq4$ dropout selection criteria (accomodating slightly redder sources) at our $i_{\rm AB}<26.2$ ($5\sigma$) depth, we find four LBGs that are likely new members of an already highly overdense region: one is unambiguously the counterpart of an ALMA SMG ($M$); one is only 0.8\,arcsec offset from another ALMA SMG ($J$); one has a candidate \CII\ line detection in our deep ALMA data; and the final one is detected in Ly$\alpha$ by VLT/MUSE (Apostolovski in prep.). While this represents a substantial local overdensity (perhaps 8-17 times the background level within a 30\,arcsec-diameter region), the small number of galaxies would not be a statistically significant overdensity in a large survey. The $i_{\rm AB}<25$ LBG overdensity in a 1.8\,arcmin-radius aperture, matching the \cite{Toshikawa18} proto-cluster search criteria, is only 3$\sigma$ significant for SPT2349-56.

\item For the nine SMGs detected by IRAC or in $K_{\rm s}$, we use their multi-band imaging 
to study their  properties 
and estimate the stellar masses using \textsc{cigale} fitting, finding $M^*$ to range between $(0.4{-}3.2)\times10^{11}$\,\msun. 

\item Source $C$ appears to be brighter and to have a substantially larger stellar mass than any of the other SPT2349-56 sources, and is amongst the most massive $z>3.5$ SMGs from the ALESS field survey. It may be the stellar seed of a rapidly forming BCG galaxy.

\item The highest SFR SMGs in SPT2349-56 have a similar range of \mstar\ to the ALESS SMGs at $z>3.5$, and lie within the scatter of the main sequence of star formation at these redshifts. However, our lowest SFR SMGs have stellar masses consistent with the high \mstar\ values of the other SPT2349-56 sources as well as the ALESS SMGs 
and may represent rapid build-up of stellar mass and a subsequent early quenching of massive galaxies in the dense proto-cluster core.

\item The cumulative stellar mass for the SPT2349-56 core is at least $(12.2\pm2.8)\times10^{11}$\,\msun. Since these galaxies appear destined to merge on a short timescale \citep{Rennehan}, this stellar mass is already comparable to that of a BCG in a ${>}\,10^{15}$\,\msun\ galaxy cluster at $z<1$. The combined stars and gas in the SPT2349-56 core represent a large fraction (more than $\sim19$ per cent) of the virial mass estimate, and this may suggest that there is not yet an established hot ICM at this early epoch.

\end{itemize}

\section*{Acknowledgements}
%
We thank Mischa Schirmer for help with the Gemini imaging reductions. We further thank the referee for a constructive and helpful report.
This paper was based on observations obtained at the international Gemini Observatory, a programme of NSF’s OIR Lab, which is managed by the Association of Universities for Research in Astronomy (AURA) under a cooperative agreement with the National Science Foundation on behalf of the Gemini Observatory partnership: the National Science Foundation (United States), National Research Council (Canada), Agencia Nacional de Investigación y Desarrollo(Chile), Ministerio de Ciencia, Tecnología e Innovación (Argentina), Ministério da Ciência, Tecnologia, Inovações e Comunicações (Brazil), and Korea Astronomy and Space Science Institute (Republic of Korea), Program GS-2017B-Q-7 (PI: Chapman).
Also based on observations collected at the European Organisation for Astronomical Research in the Southern Hemisphere under ESO programme 092.A-0503(A).
This work is based in part on observations made with the {\it Spitzer Space Telescope}, which was operated by the Jet Propulsion Laboratory, California Institute of Technology under a contract with NASA.
This paper makes use of the following ALMA data: ADS/JAO.ALMA\#2017.1.00273.S; and ADS/JAO.ALMA\#2018.1.00058.S. 
ALMA is a partnership of ESO (representing its member states), NSF (USA) and NINS (Japan), together with NRC (Canada), MOST and ASIAA (Taiwan), and KASI (Republic of Korea), in cooperation with the Republic of Chile. The Joint ALMA Observatory is operated by ESO, AUI/NRAO and NAOJ.
The National Radio Astronomy Observatory is a facility of the National Science Foundation operated under cooperative agreement by Associated Universities, Inc.
The SPT is supported by the NSF through grant OPP-1852617.
K.M.R.\ and S.C.C.\ acknowledge support from  NSERC, CFI, and the Killam Trust. 
A.B.\ acknowledges NSERC.
D.P.M., J.D.V., K.C.L., K.P.\ and S.J.\ acknowledge support from the US NSF under grants AST-1715213 and AST-1716127.
S.J.\ and K.C.L acknowledge support from the US NSF NRAO under grants SOSPA5-001 and SOSPA4-007, respectively.
J.D.V.\ acknowledges support from an A. P. Sloan Foundation Fellowship.

\section*{Data availability}
The data underlying this article will be shared on reasonable request to the corresponding author.

\bibliographystyle{mnras}
\bibliography{SPT2349optical}

\begin{thebibliography}{}
\makeatletter
\relax
\def\mn@urlcharsother{\let\do\@makeother \do\$\do\&\do\#\do\^\do\_\do\%\do\~}
\def\mn@doi{\begingroup\mn@urlcharsother \@ifnextchar [ {\mn@doi@}
  {\mn@doi@[]}}
\def\mn@doi@[#1]#2{\def\@tempa{#1}\ifx\@tempa\@empty \href
  {http://dx.doi.org/#2} {doi:#2}\else \href {http://dx.doi.org/#2} {#1}\fi
  \endgroup}
\def\mn@eprint#1#2{\mn@eprint@#1:#2::\@nil}
\def\mn@eprint@arXiv#1{\href {http://arxiv.org/abs/#1} {{\tt arXiv:#1}}}
\def\mn@eprint@dblp#1{\href {http://dblp.uni-trier.de/rec/bibtex/#1.xml}
  {dblp:#1}}
\def\mn@eprint@#1:#2:#3:#4\@nil{\def\@tempa {#1}\def\@tempb {#2}\def\@tempc
  {#3}\ifx \@tempc \@empty \let \@tempc \@tempb \let \@tempb \@tempa \fi \ifx
  \@tempb \@empty \def\@tempb {arXiv}\fi \@ifundefined
  {mn@eprint@\@tempb}{\@tempb:\@tempc}{\expandafter \expandafter \csname
  mn@eprint@\@tempb\endcsname \expandafter{\@tempc}}}

\bibitem[\protect\citeauthoryear{Ashby et~al.,}{Ashby et~al.}{2013}]{Ashby13}
Ashby M.~L.~N.,  et~al., 2013, \mn@doi [ApJS] {10.1088/0067-0049/209/2/22},
  \href {https://ui.adsabs.harvard.edu/abs/2013ApJS..209...22A} {209, 22}

\bibitem[\protect\citeauthoryear{Bertin \& Arnouts}{Bertin \&
  Arnouts}{1996}]{Bertin}
Bertin E.,  Arnouts S.,  1996, \mn@doi [A\&AS] {10.1051/aas:1996164}, \href
  {https://ui.adsabs.harvard.edu/abs/1996A%26AS..117..393B} {117, 393}

\bibitem[\protect\citeauthoryear{Boquien, Burgarella, Roehlly, Buat, Ciesla,
  Corre, Inoue  \& Salas}{Boquien et~al.}{2019}]{Boquien19}
Boquien M.,  Burgarella D.,  Roehlly Y.,  Buat V.,  Ciesla L.,  Corre D.,
  Inoue A.~K.,   Salas H.,  2019, \mn@doi [A\&A] {10.1051/0004-6361/201834156},
  \href {https://ui.adsabs.harvard.edu/abs/2019A&A...622A.103B} {622, A103}

\bibitem[\protect\citeauthoryear{Bothwell et~al.,}{Bothwell
  et~al.}{2013}]{Bothwell}
Bothwell M.~S.,  et~al., 2013, \mn@doi [MNRAS] {10.1093/mnras/sts562}, 429,
  3047

\bibitem[\protect\citeauthoryear{Calzetti, Armus, Bohlin, Kinney, Koornneef  \&
  Storchi-Bergmann}{Calzetti et~al.}{2000}]{Calzetti00}
Calzetti D.,  Armus L.,  Bohlin R.~C.,  Kinney A.~L.,  Koornneef J.,
  Storchi-Bergmann T.,  2000, \mn@doi [ApJ] {10.1086/308692}, \href
  {https://ui.adsabs.harvard.edu/abs/2000ApJ...533..682C} {533, 682}

\bibitem[\protect\citeauthoryear{Casey}{Casey}{2016}]{Casey16}
Casey C.~M.,  2016, \mn@doi [ApJ] {10.3847/0004-637X/824/1/36}, \href
  {https://ui.adsabs.harvard.edu/abs/2016ApJ...824...36C} {824, 36}

\bibitem[\protect\citeauthoryear{Casey, Narayanan  \& Cooray}{Casey
  et~al.}{2014}]{Casey}
Casey C.~M.,  Narayanan D.,   Cooray A.,  2014, \mn@doi [Phys. Rep.]
  {https://doi.org/10.1016/j.physrep.2014.02.009}, 541, 45

\bibitem[\protect\citeauthoryear{Chabrier}{Chabrier}{2003}]{Chabrier}
Chabrier G.,  2003, \mn@doi [PASP] {10.1086/376392}, \href
  {https://ui.adsabs.harvard.edu/abs/2003PASP..115..763C} {115, 763}

\bibitem[\protect\citeauthoryear{Chapman, Blain, Ivison  \& Smail}{Chapman
  et~al.}{2003}]{Chapman03N}
Chapman S.~C.,  Blain A.~W.,  Ivison R.~J.,   Smail I.~R.,  2003, \mn@doi
  [Nature] {10.1038/nature01540}, 422, 695

\bibitem[\protect\citeauthoryear{Chapman, Blain, Smail  \& Ivison}{Chapman
  et~al.}{2005}]{Chapman05}
Chapman S.~C.,  Blain A.~W.,  Smail I.,   Ivison R.~J.,  2005, \mn@doi [ApJ]
  {10.1086/428082}, \href {http://adsabs.harvard.edu/abs/2005ApJ...622..772C}
  {622, 772}

\bibitem[\protect\citeauthoryear{Chapman, Blain, Ibata, Ivison, Smail  \&
  Morrison}{Chapman et~al.}{2009}]{Chapman09}
Chapman S.~C.,  Blain A.,  Ibata R.,  Ivison R.~J.,  Smail I.,   Morrison G.,
  2009, \mn@doi [ApJ] {10.1088/0004-637X/691/1/560}, \href
  {https://ui.adsabs.harvard.edu/abs/2009ApJ...691..560C} {691, 560}

\bibitem[\protect\citeauthoryear{Chiang, Overzier  \& Gebhardt}{Chiang
  et~al.}{2013}]{Chiang13}
Chiang Y.-K.,  Overzier R.,   Gebhardt K.,  2013, \mn@doi [ApJ]
  {10.1088/0004-637X/779/2/127}, \href
  {https://ui.adsabs.harvard.edu/abs/2013ApJ...779..127C} {779, 127}

\bibitem[\protect\citeauthoryear{Chiang, Overzier, Gebhardt  \&
  Henriques}{Chiang et~al.}{2017}]{Chiang17}
Chiang Y.-K.,  Overzier R.~A.,  Gebhardt K.,   Henriques B.,  2017, \mn@doi
  [ApJL] {10.3847/2041-8213/aa7e7b}, \href
  {https://ui.adsabs.harvard.edu/abs/2017ApJ...844L..23C} {844, L23}

\bibitem[\protect\citeauthoryear{Ciesla et~al.,}{Ciesla
  et~al.}{2015}]{Ciesla15}
Ciesla L.,  et~al., 2015, \mn@doi [A\&A] {10.1051/0004-6361/201425252}, \href
  {https://ui.adsabs.harvard.edu/abs/2015A&A...576A..10C} {576, A10}

\bibitem[\protect\citeauthoryear{Collins et~al.,}{Collins
  et~al.}{2009}]{Collins09}
Collins C.~A.,  et~al., 2009, \mn@doi [Nature] {10.1038/nature07865}, \href
  {https://ui.adsabs.harvard.edu/abs/2009Natur.458..603C} {458, 603}

\bibitem[\protect\citeauthoryear{Cooke, Kartaltepe, Tyler, Darvish, Casey,
  Le~F{\`e}vre, Salvato  \& Scoville}{Cooke et~al.}{2019}]{Cooke19}
Cooke K.~C.,  Kartaltepe J.~S.,  Tyler K.~D.,  Darvish B.,  Casey C.~M.,
  Le~F{\`e}vre O.,  Salvato M.,   Scoville N.,  2019, \mn@doi [ApJ]
  {10.3847/1538-4357/ab30c9}, \href
  {https://ui.adsabs.harvard.edu/abs/2019ApJ...881..150C} {881, 150}

\bibitem[\protect\citeauthoryear{Cowley, B{\'e}thermin, Lagos, Lacey, Baugh  \&
  Cole}{Cowley et~al.}{2017}]{Cowley}
Cowley W.~I.,  B{\'e}thermin M.,  Lagos C. d.~P.,  Lacey C.~G.,  Baugh C.~M.,
  Cole S.,  2017, \mn@doi [MNRAS] {10.1093/mnras/stx165}, \href
  {https://ui.adsabs.harvard.edu/abs/2017MNRAS.467.1231C} {467, 1231}

\bibitem[\protect\citeauthoryear{Cucciati et~al.,}{Cucciati
  et~al.}{2018}]{Cucciati18}
Cucciati O.,  et~al., 2018, \mn@doi [A\&A] {10.1051/0004-6361/201833655}, \href
  {https://ui.adsabs.harvard.edu/abs/2018A&A...619A..49C} {619, A49}

\bibitem[\protect\citeauthoryear{Damjanov et~al.,}{Damjanov
  et~al.}{2011}]{Damjanov}
Damjanov I.,  et~al., 2011, \mn@doi [ApJL] {10.1088/2041-8205/739/2/L44}, \href
  {https://ui.adsabs.harvard.edu/abs/2011ApJ...739L..44D} {739, L44}

\bibitem[\protect\citeauthoryear{Draine \& Li}{Draine \& Li}{2007}]{Draine07}
Draine B.~T.,  Li A.,  2007, \mn@doi [ApJ] {10.1086/511055}, \href
  {https://ui.adsabs.harvard.edu/abs/2007ApJ...657..810D} {657, 810}

\bibitem[\protect\citeauthoryear{Draine et~al.,}{Draine
  et~al.}{2014}]{Draine14}
Draine B.~T.,  et~al., 2014, \mn@doi [ApJ] {10.1088/0004-637X/780/2/172}, \href
  {https://ui.adsabs.harvard.edu/abs/2014ApJ...780..172D} {780, 172}

\bibitem[\protect\citeauthoryear{Eikenberry et~al.,}{Eikenberry
  et~al.}{2004}]{Eikenberry}
Eikenberry S.~S.,  et~al., 2004, in {Moorwood} A. F.~M.,  {Iye} M.,  eds,
  Society of Photo-Optical Instrumentation Engineers (SPIE) Conference Series
  Vol. 5492, Proc SPIE. pp 1196--1207, \mn@doi{10.1117/12.549796}

\bibitem[\protect\citeauthoryear{Engel et~al.,}{Engel et~al.}{2010}]{Engel10}
Engel H.,  et~al., 2010, \mn@doi [ApJ] {10.1088/0004-637X/724/1/233}, \href
  {https://ui.adsabs.harvard.edu/abs/2010ApJ...724..233E} {724, 233}

\bibitem[\protect\citeauthoryear{Erb, Steidel, Shapley, Pettini, Reddy  \&
  Adelberger}{Erb et~al.}{2006}]{Erb06}
Erb D.~K.,  Steidel C.~C.,  Shapley A.~E.,  Pettini M.,  Reddy N.~A.,
  Adelberger K.~L.,  2006, \mn@doi [ApJ] {10.1086/504891}, \href
  {https://ui.adsabs.harvard.edu/abs/2006ApJ...646..107E} {646, 107}

\bibitem[\protect\citeauthoryear{Everett et~al.,}{Everett
  et~al.}{2020}]{Everett20}
Everett W.~B.,  et~al., 2020, arXiv e-prints, \href
  {https://ui.adsabs.harvard.edu/abs/2020arXiv200303431E} {p. arXiv:2003.03431}

\bibitem[\protect\citeauthoryear{Fazio et~al.,}{Fazio et~al.}{2004}]{Fazio04}
Fazio G.,  et~al., 2004, \mn@doi [ApJS] {10.1086/422843}, 154, 10

\bibitem[\protect\citeauthoryear{Fritz, Franceschini  \& Hatziminaoglou}{Fritz
  et~al.}{2006}]{Fritz06}
Fritz J.,  Franceschini A.,   Hatziminaoglou E.,  2006, \mn@doi [MNRAS]
  {10.1111/j.1365-2966.2006.09866.x}, \href
  {https://ui.adsabs.harvard.edu/abs/2006MNRAS.366..767F} {366, 767}

\bibitem[\protect\citeauthoryear{Hainline, Blain, Smail, Alexander, Armus,
  Chapman  \& Ivison}{Hainline et~al.}{2011}]{Hainline11}
Hainline L.~J.,  Blain A.~W.,  Smail I.,  Alexander D.~M.,  Armus L.,  Chapman
  S.~C.,   Ivison R.~J.,  2011, \mn@doi [ApJ] {10.1088/0004-637x/740/2/96},
  740, 96

\bibitem[\protect\citeauthoryear{Harikane et~al.,}{Harikane
  et~al.}{2019}]{Harikane19}
Harikane Y.,  et~al., 2019, \mn@doi [ApJ] {10.3847/1538-4357/ab2cd5}, \href
  {https://ui.adsabs.harvard.edu/abs/2019ApJ...883..142H} {883, 142}

\bibitem[\protect\citeauthoryear{Higuchi et~al.,}{Higuchi
  et~al.}{2019}]{Higuchi19}
Higuchi R.,  et~al., 2019, \mn@doi [ApJ] {10.3847/1538-4357/ab2192}, \href
  {https://ui.adsabs.harvard.edu/abs/2019ApJ...879...28H} {879, 28}

\bibitem[\protect\citeauthoryear{Hill et~al.,}{Hill et~al.}{2020}]{Hill20}
Hill R.,  et~al., 2020, \mn@doi [MNRAS] {10.1093/mnras/staa1275}, \href
  {https://ui.adsabs.harvard.edu/abs/2020MNRAS.495.3124H} {495, 3124}

\bibitem[\protect\citeauthoryear{Hilton et~al.,}{Hilton
  et~al.}{2013}]{Hilton13}
Hilton M.,  et~al., 2013, \mn@doi [MNRAS] {10.1093/mnras/stt1535}, \href
  {https://ui.adsabs.harvard.edu/abs/2013MNRAS.435.3469H} {435, 3469}

\bibitem[\protect\citeauthoryear{Hook, J{\o}rgensen, Allington-Smith, Davies,
  Metcalfe, Murowinski  \& Crampton}{Hook et~al.}{2004}]{Hook}
Hook I.~M.,  J{\o}rgensen I.,  Allington-Smith J.~R.,  Davies R.~L.,  Metcalfe
  N.,  Murowinski R.~G.,   Crampton D.,  2004, \mn@doi [PASP] {10.1086/383624},
  \href {https://ui.adsabs.harvard.edu/abs/2004PASP..116..425H} {116, 425}

\bibitem[\protect\citeauthoryear{Inoue}{Inoue}{2011}]{Inoue11}
Inoue A.~K.,  2011, \mn@doi [MNRAS] {10.1111/j.1365-2966.2011.18906.x}, \href
  {https://ui.adsabs.harvard.edu/abs/2011MNRAS.415.2920I} {415, 2920}

\bibitem[\protect\citeauthoryear{Ivison, Biggs, Bremer, Arumugam  \&
  Dunne}{Ivison et~al.}{2020}]{Ivison20}
Ivison R.~J.,  Biggs A.~D.,  Bremer M.,  Arumugam V.,   Dunne L.,  2020, arXiv
  e-prints, \href {https://ui.adsabs.harvard.edu/abs/2020arXiv200610753I} {p.
  arXiv:2006.10753}

\bibitem[\protect\citeauthoryear{Kurk, Pentericci, R{\"o}ttgering  \&
  Miley}{Kurk et~al.}{2004}]{Kurk04}
Kurk J.~D.,  Pentericci L.,  R{\"o}ttgering H.~J.~A.,   Miley G.~K.,  2004,
  \mn@doi [A\&A] {10.1051/0004-6361:20040075}, \href
  {https://ui.adsabs.harvard.edu/abs/2004A&A...428..793K} {428, 793}

\bibitem[\protect\citeauthoryear{Lacaille et~al.,}{Lacaille
  et~al.}{2019}]{Lacaille19}
Lacaille K.~M.,  et~al., 2019, \mn@doi [MNRAS] {10.1093/mnras/stz1742}, \href
  {https://ui.adsabs.harvard.edu/abs/2019MNRAS.488.1790L} {488, 1790}

\bibitem[\protect\citeauthoryear{Le~F{\`e}vre et~al.,}{Le~F{\`e}vre
  et~al.}{2015}]{Fevre15}
Le~F{\`e}vre O.,  et~al., 2015, \mn@doi [A\&A] {10.1051/0004-6361/201423829},
  \href {https://ui.adsabs.harvard.edu/abs/2015A&A...576A..79L} {576, A79}

\bibitem[\protect\citeauthoryear{Lemaux et~al.,}{Lemaux
  et~al.}{2018}]{Lemaux18}
Lemaux B.~C.,  et~al., 2018, \mn@doi [A\&A] {10.1051/0004-6361/201730870},
  \href {https://ui.adsabs.harvard.edu/abs/2018A&A...615A..77L} {615, A77}

\bibitem[\protect\citeauthoryear{Lewis et~al.,}{Lewis et~al.}{2018}]{Lewis18}
Lewis A.~J.~R.,  et~al., 2018, \mn@doi [ApJ] {10.3847/1538-4357/aacc25}, \href
  {https://ui.adsabs.harvard.edu/abs/2018ApJ...862...96L} {862, 96}

\bibitem[\protect\citeauthoryear{Long et~al.,}{Long et~al.}{2020}]{Long20}
Long A.~S.,  et~al., 2020, arXiv e-prints, \href
  {https://ui.adsabs.harvard.edu/abs/2020arXiv200313694L} {p. arXiv:2003.13694}

\bibitem[\protect\citeauthoryear{Ma et~al.,}{Ma et~al.}{2015}]{Ma15}
Ma J.,  et~al., 2015, \mn@doi [ApJ] {10.1088/0004-637x/812/1/88}, 812, 88

\bibitem[\protect\citeauthoryear{Micha{\l}owski, Dunlop, Cirasuolo, Hjorth,
  Hayward  \& Watson}{Micha{\l}owski et~al.}{2012}]{Michalowski12}
Micha{\l}owski M.~J.,  Dunlop J.~S.,  Cirasuolo M.,  Hjorth J.,  Hayward C.~C.,
    Watson D.,  2012, \mn@doi [A\&A] {10.1051/0004-6361/201016308}, \href
  {http://adsabs.harvard.edu/abs/2012A%26A...541A..85M} {541, A85}

\bibitem[\protect\citeauthoryear{Miller, Hayward, Chapman  \& Behroozi}{Miller
  et~al.}{2015}]{Miller15}
Miller T.~B.,  Hayward C.~C.,  Chapman S.~C.,   Behroozi P.~S.,  2015, \mn@doi
  [MNRAS] {10.1093/mnras/stv1267}, \href
  {https://ui.adsabs.harvard.edu/abs/2015MNRAS.452..878M} {452, 878}

\bibitem[\protect\citeauthoryear{Miller et~al.,}{Miller et~al.}{2018}]{Miller}
Miller T.~B.,  et~al., 2018, \mn@doi [Nature] {10.1038/s41586-018-0025-2}, 556,
  469

\bibitem[\protect\citeauthoryear{Mocanu et~al.,}{Mocanu et~al.}{2013}]{Mocanu}
Mocanu L.~M.,  et~al., 2013, \mn@doi [ApJ] {10.1088/0004-637x/779/1/61}, 779,
  61

\bibitem[\protect\citeauthoryear{Modigliani et~al.,}{Modigliani
  et~al.}{2010}]{Modigliani}
Modigliani A.,  et~al., 2010, \mn@doi [Proceedings of SPIE]
  {10.1117/12.857211}, 7737, 56

\bibitem[\protect\citeauthoryear{Muldrew, Hatch  \& Cooke}{Muldrew
  et~al.}{2015}]{Muldrew15}
Muldrew S.~I.,  Hatch N.~A.,   Cooke E.~A.,  2015, \mn@doi [MNRAS]
  {10.1093/mnras/stv1449}, \href
  {https://ui.adsabs.harvard.edu/abs/2015MNRAS.452.2528M} {452, 2528}

\bibitem[\protect\citeauthoryear{Narayanan et~al.,}{Narayanan
  et~al.}{2010}]{Narayanan10}
Narayanan D.,  et~al., 2010, \mn@doi [MNRAS]
  {10.1111/j.1365-2966.2010.16997.x}, \href
  {https://ui.adsabs.harvard.edu/abs/2010MNRAS.407.1701N} {407, 1701}

\bibitem[\protect\citeauthoryear{Noeske et~al.,}{Noeske et~al.}{2007}]{Noeske}
Noeske K.~G.,  et~al., 2007, \mn@doi [ApJ] {10.1086/517926}, \href
  {https://ui.adsabs.harvard.edu/abs/2007ApJ...660L..43N} {660, L43}

\bibitem[\protect\citeauthoryear{Noll, Burgarella, Giovannoli, Buat, Marcillac
  \& Mu\~noz Mateos}{Noll et~al.}{2009}]{Noll09}
Noll S.,  Burgarella D.,  Giovannoli E.,  Buat V.,  Marcillac D.,   Mu\~noz
  Mateos J.~C.,  2009, \mn@doi [A\&A] {10.1051/0004-6361/200912497}, 507, 1793

\bibitem[\protect\citeauthoryear{Oteo et~al.,}{Oteo et~al.}{2018}]{Oteo18}
Oteo I.,  et~al., 2018, \mn@doi [ApJ] {10.3847/1538-4357/aaa1f1}, \href
  {https://ui.adsabs.harvard.edu/abs/2018ApJ...856...72O} {856, 72}

\bibitem[\protect\citeauthoryear{Overzier}{Overzier}{2016}]{Overzier}
Overzier R.~A.,  2016, \mn@doi [A\&AR] {10.1007/s00159-016-0100-3}, 24, 14

\bibitem[\protect\citeauthoryear{Pentericci et~al.,}{Pentericci
  et~al.}{2000}]{Pentericci00}
Pentericci L.,  et~al., 2000, A\&A, \href
  {https://ui.adsabs.harvard.edu/abs/2000A&A...361L..25P} {361, L25}

\bibitem[\protect\citeauthoryear{Ragone-Figueroa, Granato, Ferraro, Murante,
  Biffi, Borgani, Planelles  \& Rasia}{Ragone-Figueroa et~al.}{2018}]{Ragone18}
Ragone-Figueroa C.,  Granato G.~L.,  Ferraro M.~E.,  Murante G.,  Biffi V.,
  Borgani S.,  Planelles S.,   Rasia E.,  2018, \mn@doi [MNRAS]
  {10.1093/mnras/sty1639}, \href
  {https://ui.adsabs.harvard.edu/abs/2018MNRAS.479.1125R} {479, 1125}

\bibitem[\protect\citeauthoryear{Rennehan, Babul, Hayward, Bottrell, Hani  \&
  Chapman}{Rennehan et~al.}{2020}]{Rennehan}
Rennehan D.,  Babul A.,  Hayward C.~C.,  Bottrell C.,  Hani M.~H.,   Chapman
  S.~C.,  2020, \mn@doi [MNRAS] {10.1093/mnras/staa541}, \href
  {https://ui.adsabs.harvard.edu/abs/2020MNRAS.493.4607R} {493, 4607}

\bibitem[\protect\citeauthoryear{Reuter et~al.,}{Reuter
  et~al.}{2020}]{Reuter20}
Reuter C.,  et~al., 2020, \mn@doi [ApJ] {10.3847/1538-4357/abb599}, \href
  {https://ui.adsabs.harvard.edu/abs/2020ApJ...902...78R} {902, 78}

\bibitem[\protect\citeauthoryear{Robertson et~al.,}{Robertson
  et~al.}{2019}]{Robertson19}
Robertson B.~E.,  et~al., 2019, \mn@doi [Nature Rev Phys]
  {10.1038/s42254-019-0067-x}, \href
  {https://ui.adsabs.harvard.edu/abs/2019NatRP...1..450R} {1, 450}

\bibitem[\protect\citeauthoryear{Santini et~al.,}{Santini
  et~al.}{2017}]{Santini}
Santini P.,  et~al., 2017, \mn@doi [ApJ] {10.3847/1538-4357/aa8874}, \href
  {https://ui.adsabs.harvard.edu/abs/2017ApJ...847...76S} {847, 76}

\bibitem[\protect\citeauthoryear{Schaerer et~al.,}{Schaerer
  et~al.}{2020}]{Schaerer}
Schaerer D.,  et~al., 2020, arXiv e-prints, \href
  {https://ui.adsabs.harvard.edu/abs/2020arXiv200200979S} {p. arXiv:2002.00979}

\bibitem[\protect\citeauthoryear{Serra, Amblard, Temi, Burgarella, Giovannoli,
  Buat, Noll  \& Im}{Serra et~al.}{2011}]{Serra11}
Serra P.,  Amblard A.,  Temi P.,  Burgarella D.,  Giovannoli E.,  Buat V.,
  Noll S.,   Im S.,  2011, \mn@doi [ApJ] {10.1088/0004-637X/740/1/22}, \href
  {https://ui.adsabs.harvard.edu/abs/2011ApJ...740...22S} {740, 22}

\bibitem[\protect\citeauthoryear{Shapley, Steidel, Erb, Reddy, Adelberger,
  Pettini, Barmby  \& Huang}{Shapley et~al.}{2005}]{Shapley05}
Shapley A.~E.,  Steidel C.~C.,  Erb D.~K.,  Reddy N.~A.,  Adelberger K.~L.,
  Pettini M.,  Barmby P.,   Huang J.,  2005, \mn@doi [ApJ] {10.1086/429990},
  \href {https://ui.adsabs.harvard.edu/abs/2005ApJ...626..698S} {626, 698}

\bibitem[\protect\citeauthoryear{Shimasaku et~al.,}{Shimasaku
  et~al.}{2003}]{Shimasaku03}
Shimasaku K.,  et~al., 2003, \mn@doi [The Astrophysical Journal]
  {10.1086/374880}, 586, L111

\bibitem[\protect\citeauthoryear{Simpson et~al.,}{Simpson
  et~al.}{2015}]{Simpson15}
Simpson J.~M.,  et~al., 2015, \mn@doi [ApJ] {10.1088/0004-637X/799/1/81}, \href
  {https://ui.adsabs.harvard.edu/abs/2015ApJ...799...81S} {799, 81}

\bibitem[\protect\citeauthoryear{Siringo et~al.,}{Siringo et~al.}{2009}]{APEX}
Siringo G.,  et~al., 2009, \mn@doi [A\&A] {10.1051/0004-6361/200811454}, 497,
  945

\bibitem[\protect\citeauthoryear{Soko{\l}owska, Mayer, Babul, Madau  \&
  Shen}{Soko{\l}owska et~al.}{2016}]{Sokolowska}
Soko{\l}owska A.,  Mayer L.,  Babul A.,  Madau P.,   Shen S.,  2016, \mn@doi
  [ApJ] {10.3847/0004-637X/819/1/21}, \href
  {https://ui.adsabs.harvard.edu/abs/2016ApJ...819...21S} {819, 21}

\bibitem[\protect\citeauthoryear{Speagle, Steinhardt, Capak  \&
  Silverman}{Speagle et~al.}{2014}]{Speagle14}
Speagle J.~S.,  Steinhardt C.~L.,  Capak P.~L.,   Silverman J.~D.,  2014,
  \mn@doi [ApJS] {10.1088/0067-0049/214/2/15}, \href
  {https://ui.adsabs.harvard.edu/abs/2014ApJS..214...15S} {214, 15}

\bibitem[\protect\citeauthoryear{Spilker et~al.,}{Spilker
  et~al.}{2016}]{Spilker}
Spilker J.~S.,  et~al., 2016, \mn@doi [ApJ] {10.3847/0004-637x/826/2/112}, 826,
  112

\bibitem[\protect\citeauthoryear{Stacey, Hailey-Dunsheath, Ferkinhoff, Nikola,
  Parshley, Benford, Staguhn  \& Fiolet}{Stacey et~al.}{2010}]{Stacey10}
Stacey G.~J.,  Hailey-Dunsheath S.,  Ferkinhoff C.,  Nikola T.,  Parshley
  S.~C.,  Benford D.~J.,  Staguhn J.~G.,   Fiolet N.,  2010, \mn@doi [ApJ]
  {10.1088/0004-637x/724/2/957}, 724, 957

\bibitem[\protect\citeauthoryear{Steidel, Giavalisco, Dickinson  \&
  Adelberger}{Steidel et~al.}{1996}]{Steidel96}
Steidel C.~C.,  Giavalisco M.,  Dickinson M.,   Adelberger K.~L.,  1996,
  \mn@doi [AJ] {10.1086/118019}, \href
  {https://ui.adsabs.harvard.edu/abs/1996AJ....112..352S} {112, 352}

\bibitem[\protect\citeauthoryear{Steidel, Adelberger, Dickinson, Giavalisco,
  Pettini  \& Kellogg}{Steidel et~al.}{1998}]{Steidel98}
Steidel C.~C.,  Adelberger K.~L.,  Dickinson M.,  Giavalisco M.,  Pettini M.,
  Kellogg M.,  1998, \mn@doi [ApJ] {10.1086/305073}, 492, 428

\bibitem[\protect\citeauthoryear{Steidel, Adelberger, Giavalisco, Dickinson  \&
  Pettini}{Steidel et~al.}{1999}]{Steidel99}
Steidel C.~C.,  Adelberger K.~L.,  Giavalisco M.,  Dickinson M.,   Pettini M.,
  1999, \mn@doi [ApJ] {10.1086/307363}, \href
  {https://ui.adsabs.harvard.edu/abs/1999ApJ...519....1S} {519, 1}

\bibitem[\protect\citeauthoryear{Steidel, Adelberger, Shapley, Pettini,
  Dickinson  \& Giavalisco}{Steidel et~al.}{2000}]{Steidel00}
Steidel C.~C.,  Adelberger K.~L.,  Shapley A.~E.,  Pettini M.,  Dickinson M.,
  Giavalisco M.,  2000, \mn@doi [ApJ] {10.1086/308568}, \href
  {https://ui.adsabs.harvard.edu/abs/2000ApJ...532..170S} {532, 170}

\bibitem[\protect\citeauthoryear{Steidel, Adelberger, Shapley, Erb, Reddy  \&
  Pettini}{Steidel et~al.}{2005}]{Steidel05}
Steidel C.~C.,  Adelberger K.~L.,  Shapley A.~E.,  Erb D.~K.,  Reddy N.~A.,
  Pettini M.,  2005, \mn@doi [ApJ] {10.1086/429989}, 626, 44

\bibitem[\protect\citeauthoryear{Stott, Collins, Burke, Hamilton-Morris  \&
  Smith}{Stott et~al.}{2011}]{Stott11}
Stott J.~P.,  Collins C.~A.,  Burke C.,  Hamilton-Morris V.,   Smith G.~P.,
  2011, \mn@doi [MNRAS] {10.1111/j.1365-2966.2011.18404.x}, \href
  {https://ui.adsabs.harvard.edu/abs/2011MNRAS.414..445S} {414, 445}

\bibitem[\protect\citeauthoryear{Strandet et~al.,}{Strandet
  et~al.}{2016}]{Strandet}
Strandet M.~L.,  et~al., 2016, \mn@doi [ApJ] {10.3847/0004-637x/822/2/80}, 822,
  80

\bibitem[\protect\citeauthoryear{Swinbank et~al.,}{Swinbank
  et~al.}{2013}]{Swinbank13}
Swinbank A.~M.,  et~al., 2013, \mn@doi [MNRAS] {10.1093/mnras/stt2273}, 438,
  1267

\bibitem[\protect\citeauthoryear{Tacconi et~al.,}{Tacconi
  et~al.}{2010}]{Tacconi10}
Tacconi L.~J.,  et~al., 2010, \mn@doi [Nature] {10.1038/nature08773}, \href
  {https://ui.adsabs.harvard.edu/abs/2010Natur.463..781T} {463, 781}

\bibitem[\protect\citeauthoryear{Tonini, Bernyk, Croton, Maraston  \&
  Thomas}{Tonini et~al.}{2012}]{Tonini12}
Tonini C.,  Bernyk M.,  Croton D.,  Maraston C.,   Thomas D.,  2012, \mn@doi
  [ApJ] {10.1088/0004-637X/759/1/43}, \href
  {https://ui.adsabs.harvard.edu/abs/2012ApJ...759...43T} {759, 43}

\bibitem[\protect\citeauthoryear{Toshikawa et~al.,}{Toshikawa
  et~al.}{2012}]{Toshikawa12}
Toshikawa J.,  et~al., 2012, \mn@doi [ApJ] {10.1088/0004-637x/750/2/137}, 750,
  137

\bibitem[\protect\citeauthoryear{Toshikawa et~al.,}{Toshikawa
  et~al.}{2018}]{Toshikawa18}
Toshikawa J.,  et~al., 2018, \mn@doi [PASJ] {10.1093/pasj/psx102}, \href
  {https://ui.adsabs.harvard.edu/abs/2018PASJ...70S..12T} {70, S12}

\bibitem[\protect\citeauthoryear{Umehata et~al.,}{Umehata
  et~al.}{2019}]{Umehata19}
Umehata H.,  et~al., 2019, \mn@doi [Science] {10.1126/science.aaw5949}, \href
  {https://ui.adsabs.harvard.edu/abs/2019Sci...366...97U} {366, 97}

\bibitem[\protect\citeauthoryear{Vernet et~al.,}{Vernet
  et~al.}{2011}]{Xshooter}
Vernet J.,  et~al., 2011, \mn@doi [A\&A] {10.1051/0004-6361/201117752}, \href
  {https://ui.adsabs.harvard.edu/abs/2011A%26A...536A.105V} {536, A105}

\bibitem[\protect\citeauthoryear{Vieira et~al.,}{Vieira
  et~al.}{2010}]{Vieira10}
Vieira J.~D.,  et~al., 2010, \mn@doi [ApJ] {10.1088/0004-637x/719/1/763}, 719,
  763

\bibitem[\protect\citeauthoryear{Vieira et~al.,}{Vieira
  et~al.}{2013}]{Vieira13}
Vieira J.~D.,  et~al., 2013, Nature, 495, 344 EP

\bibitem[\protect\citeauthoryear{Wei{\ss} et~al.,}{Wei{\ss}
  et~al.}{2013}]{Weiss}
Wei{\ss} A.,  et~al., 2013, \mn@doi [ApJ] {10.1088/0004-637x/767/1/88}, 767, 88

\bibitem[\protect\citeauthoryear{Wen \& Han}{Wen \& Han}{2011}]{Wen}
Wen Z.~L.,  Han J.~L.,  2011, \mn@doi [ApJ] {10.1088/0004-637X/734/1/68}, \href
  {https://ui.adsabs.harvard.edu/abs/2011ApJ...734...68W} {734, 68}

\bibitem[\protect\citeauthoryear{Werner et~al.,}{Werner
  et~al.}{2004}]{Werner04}
Werner M.~W.,  et~al., 2004, \mn@doi [ApJS] {10.1086/422992}, 154, 1

\bibitem[\protect\citeauthoryear{da Cunha et~al.,}{da~Cunha
  et~al.}{2015}]{daCunha15}
da Cunha E.,  et~al., 2015, \mn@doi [ApJ] {10.1088/0004-637x/806/1/110}, 806,
  110

\bibitem[\protect\citeauthoryear{{van der Burg} et~al.,}{{van der Burg}
  et~al.}{2013}]{vanderBurg13}
{van der Burg} R.~F.~J.,  et~al., 2013, \mn@doi [A\&A]
  {10.1051/0004-6361/201321237}, \href
  {https://ui.adsabs.harvard.edu/abs/2013A&A...557A..15V} {557, A15}

\bibitem[\protect\citeauthoryear{{van der Burg}, Hoekstra, Muzzin, Sif{\'o}n,
  Balogh  \& McGee}{{van der Burg} et~al.}{2015}]{vanderBurg15}
{van der Burg} R. F.~J.,  Hoekstra H.,  Muzzin A.,  Sif{\'o}n C.,  Balogh
  M.~L.,   McGee S.~L.,  2015, \mn@doi [A\&A] {10.1051/0004-6361/201425460},
  \href {https://ui.adsabs.harvard.edu/abs/2015A&A...577A..19V} {577, A19}

\makeatother
\end{thebibliography}

\appendix

\section*{Appendix A}\label{app:SED}
\renewcommand{\thefigure}{A\arabic{figure}}

\begin{figure*}\centering
	\includegraphics[width=0.47\textwidth]{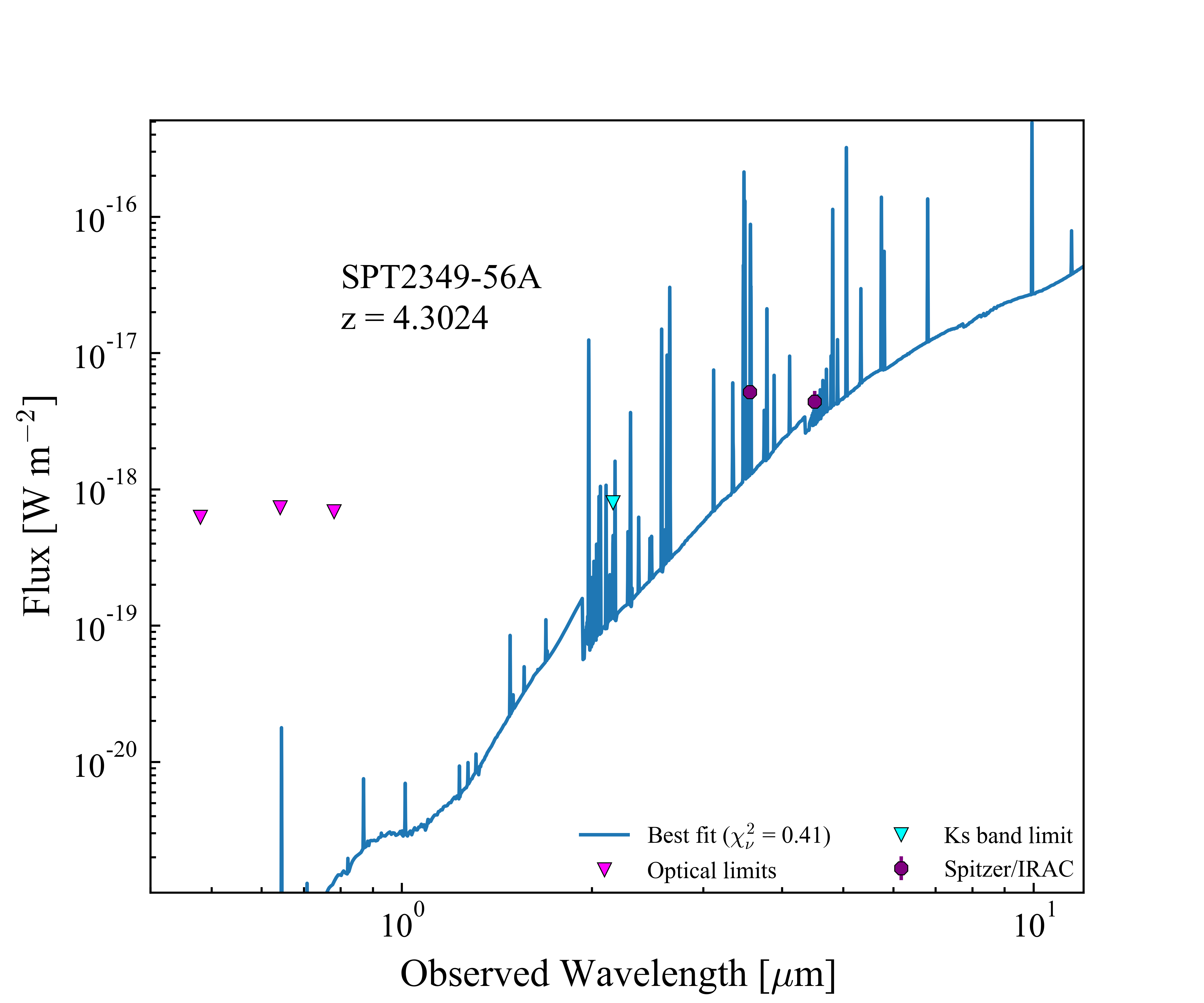}
	\includegraphics[width=0.47\textwidth]{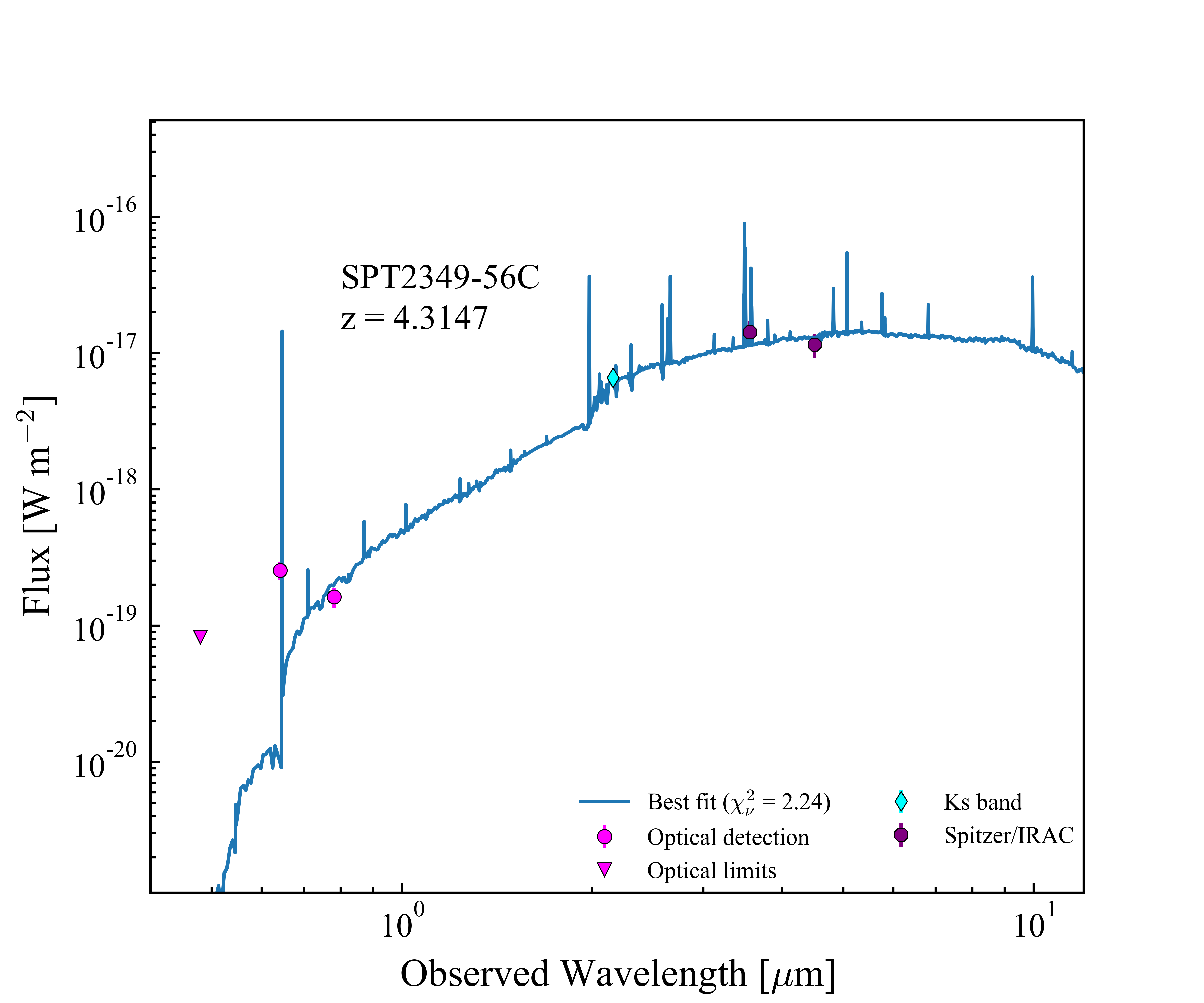}
	\includegraphics[width=0.47\textwidth]{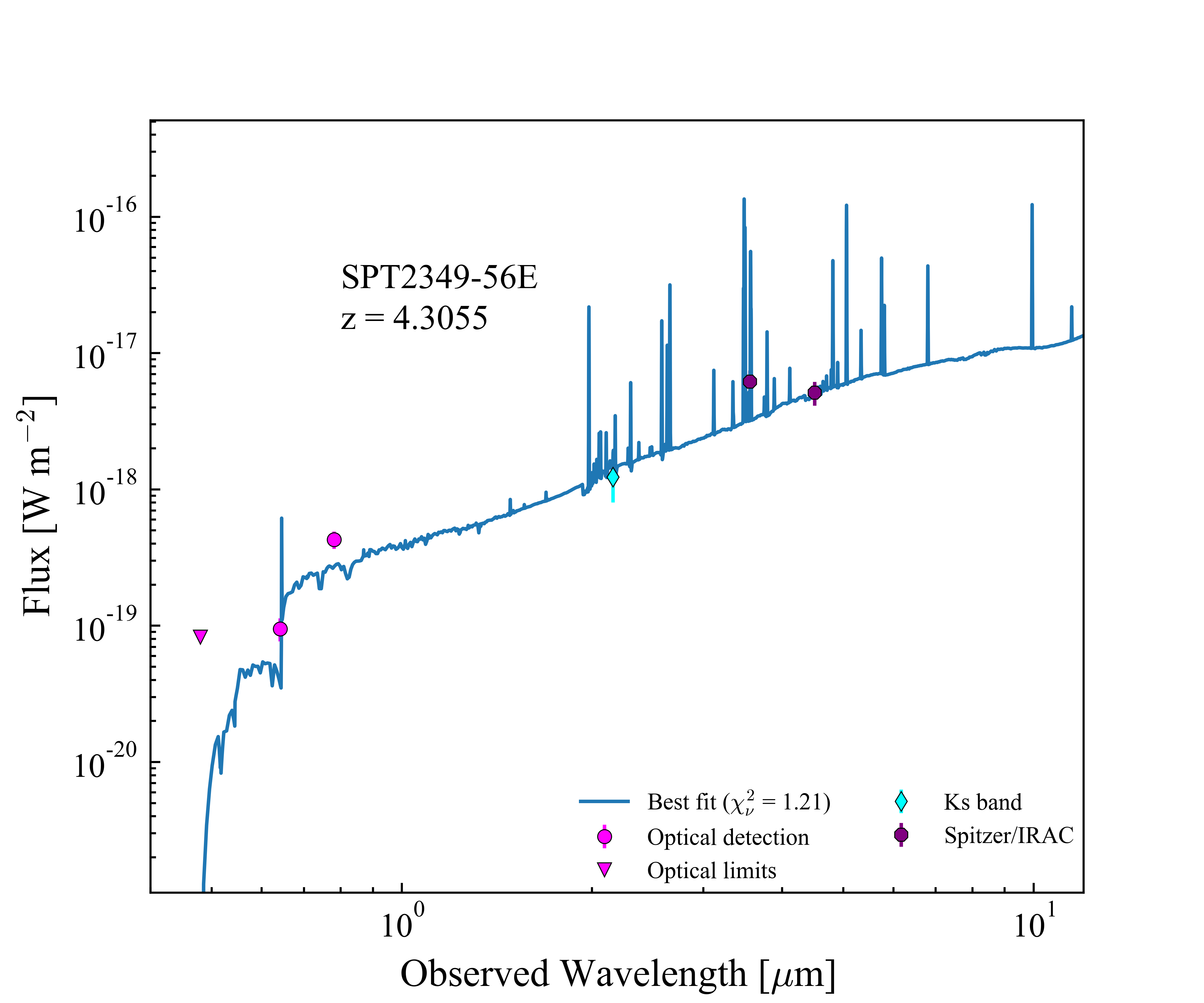}
	\includegraphics[width=0.47\textwidth]{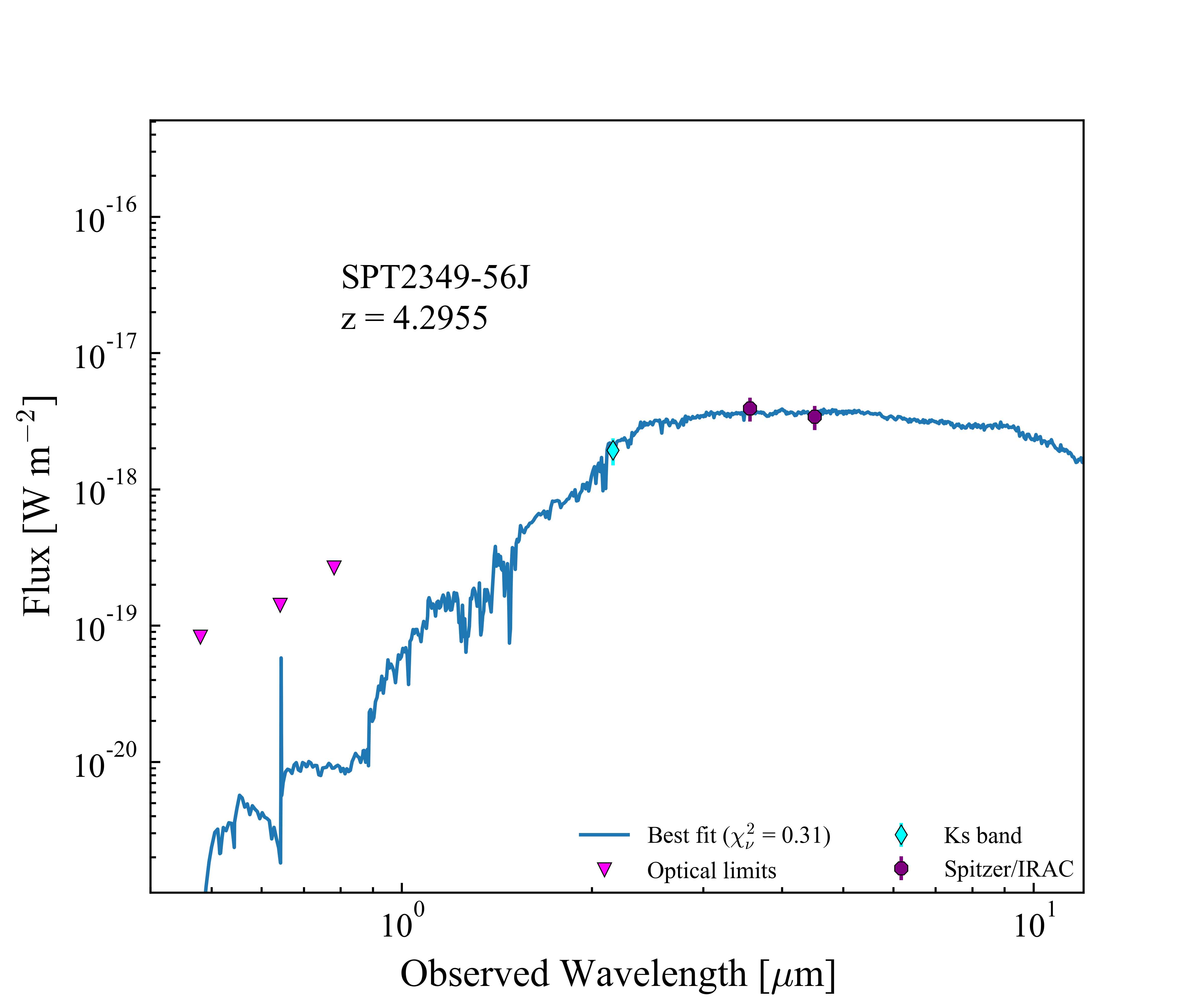}
	\includegraphics[width=0.47\textwidth]{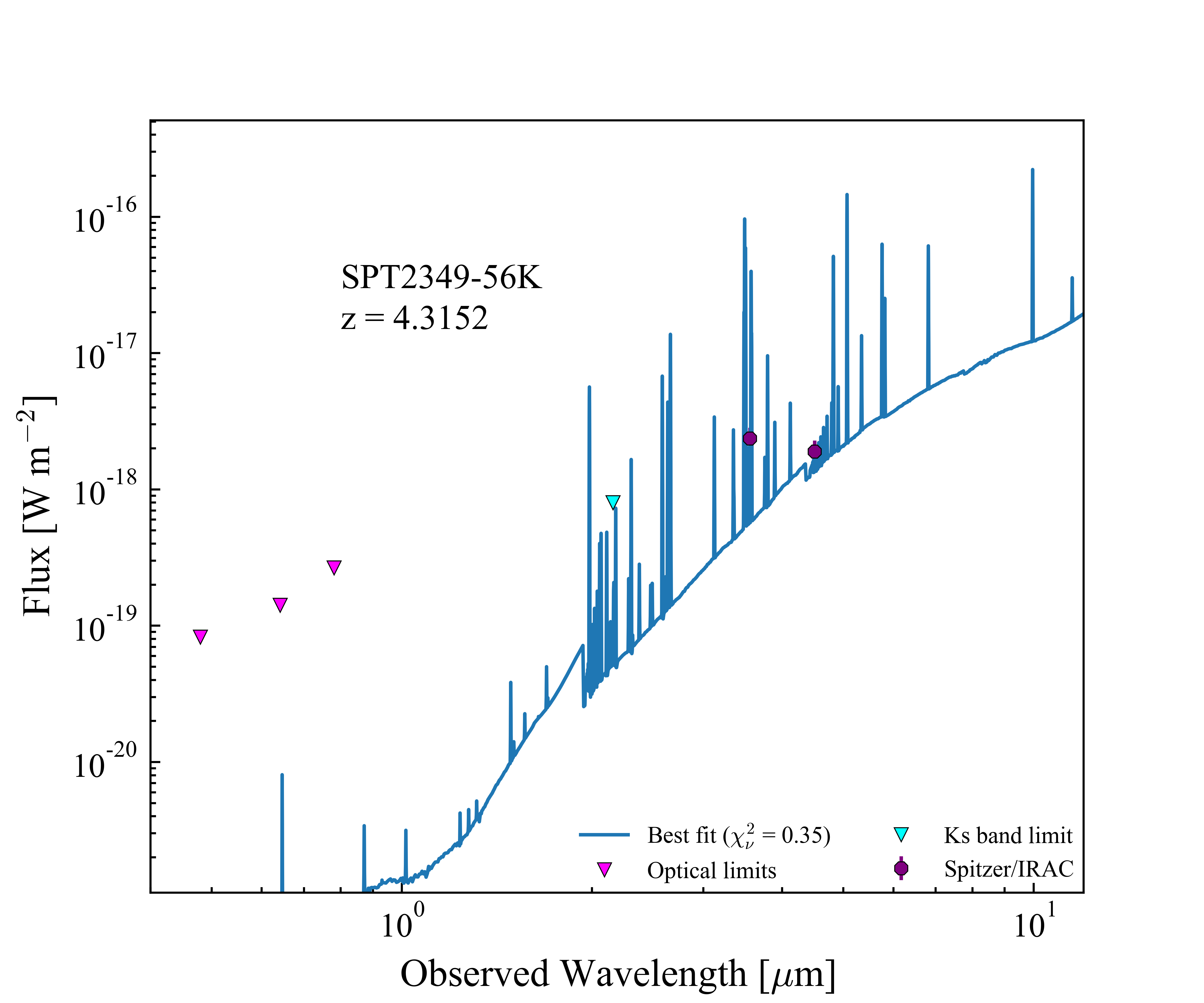}
	\includegraphics[width=0.47\textwidth]{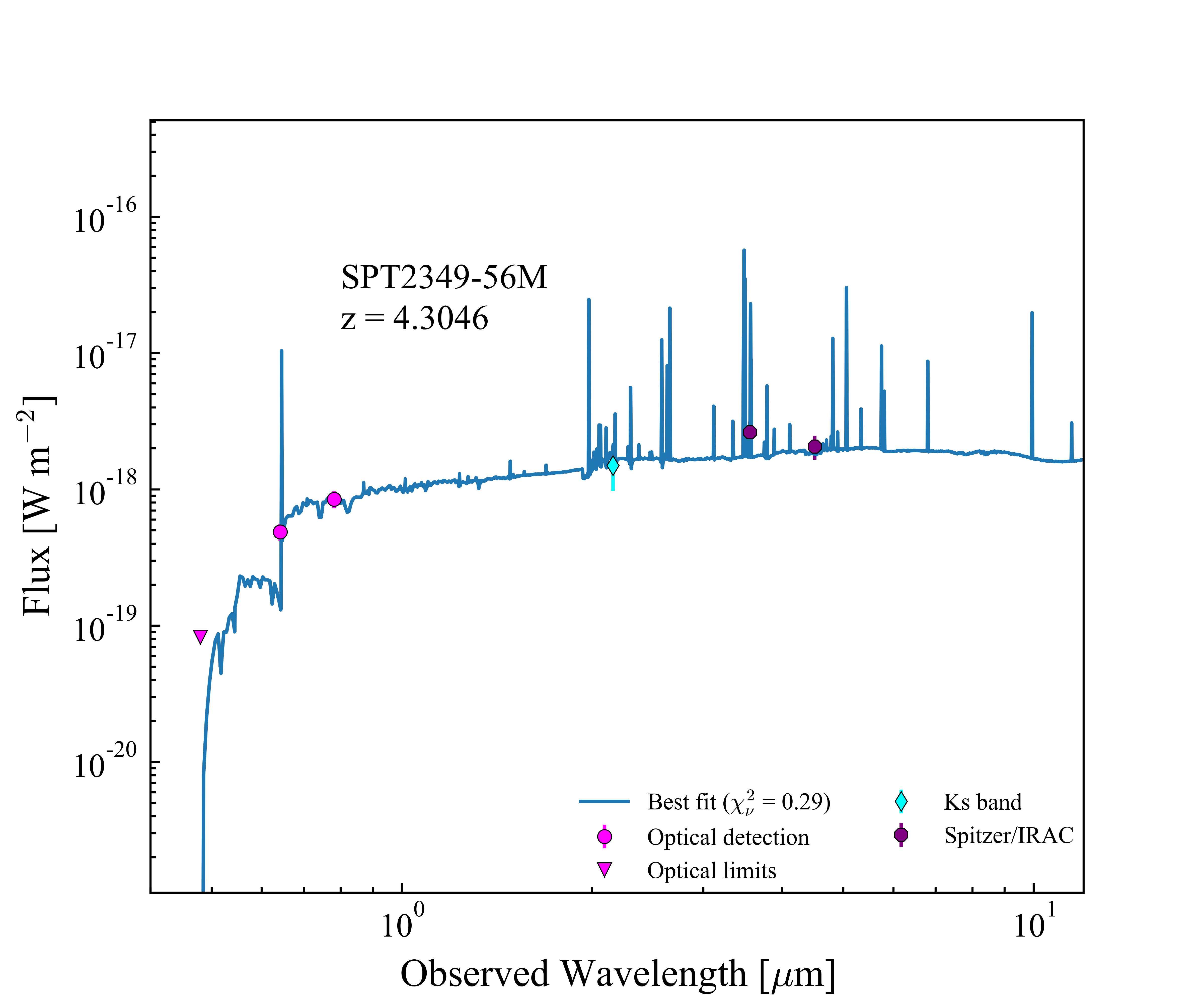}
	\caption[SED for SMGs]{Best fitting SED from \textsc{cigale} for SMGs detected at both IRAC bands.}
	\label{fig:sed_cigale}
\end{figure*}

\begin{figure*}\centering
	\includegraphics[width=0.47\textwidth]{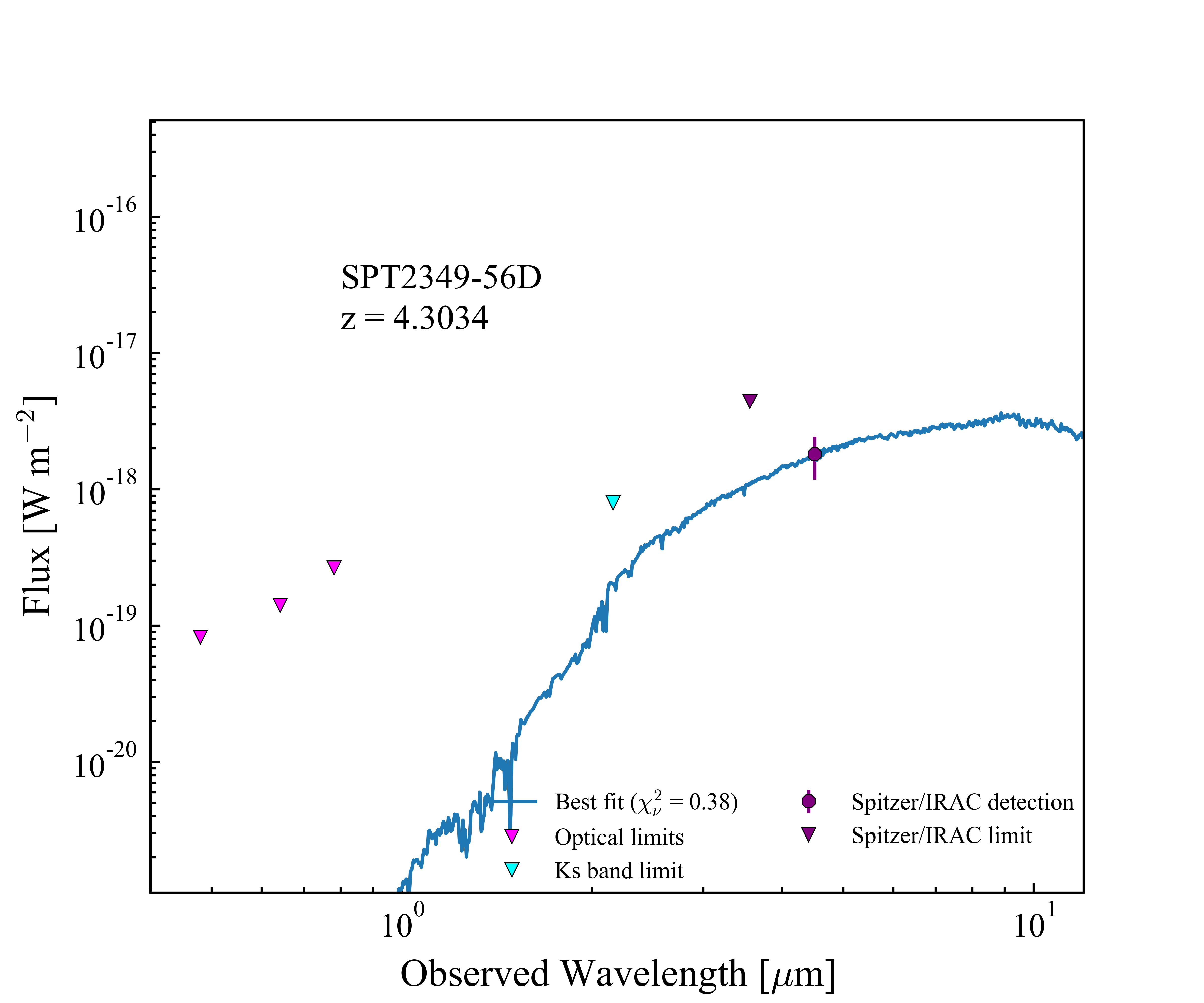}
	\includegraphics[width=0.47\textwidth]{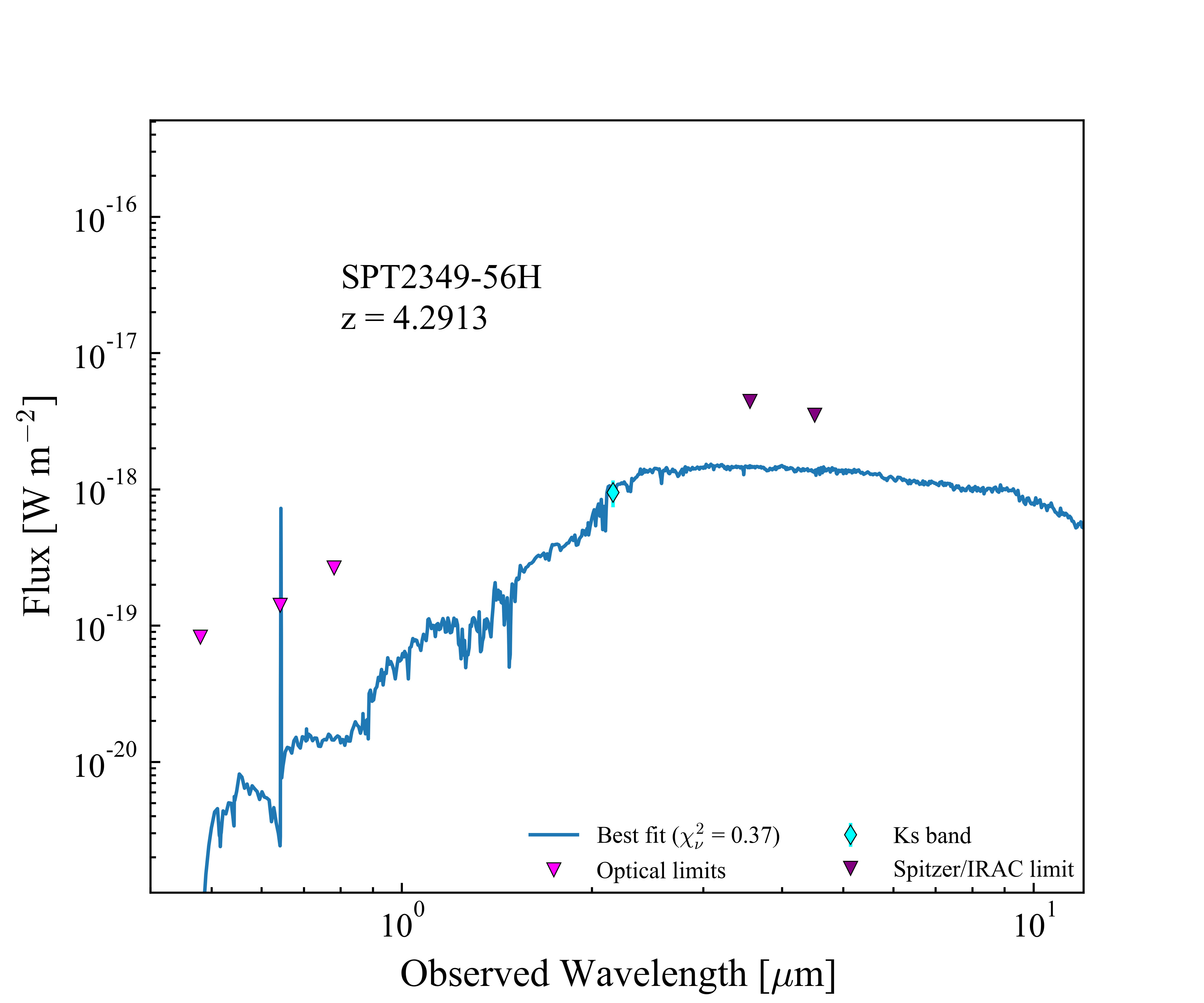}
	\includegraphics[width=0.47\textwidth]{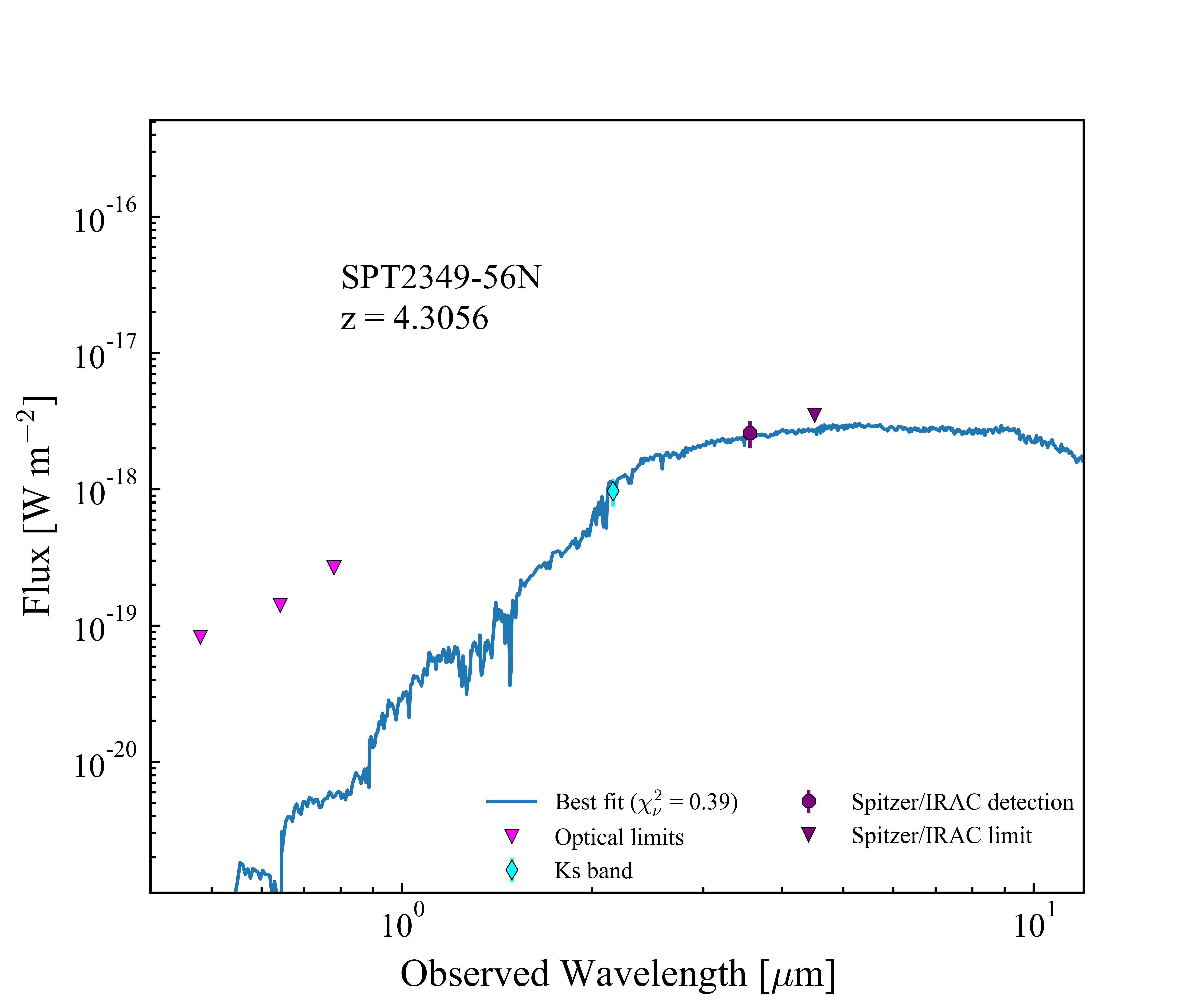}
	\caption[SED for SMGs]{Best fitting SED from \textsc{cigale} for SMGs with single band near-IR constraints. While source $N$ is detected at $3.6\,\mu$m and $K_{\rm s}$, $D$ is only detected at $4.5\,\mu$m and $H$ only at $K_{\rm s}$.}
	\label{fig:sed_cigale_single}
\end{figure*}

In the following section we include the SED fits from \textsc{cigale} for the six sources detected in both IRAC bands (see Fig. \ref{fig:sed_cigale}), followed by those detected in only one IRAC band or in $K_{\rm s}$ (see Fig. \ref{fig:sed_cigale_single}).

\section*{Appendix B}\label{app:LBG}
\renewcommand{\thefigure}{B\arabic{figure}}
\renewcommand{\thetable}{B\arabic{table}}

\begin{figure*}\centering
	\includegraphics[width=0.85\textwidth]{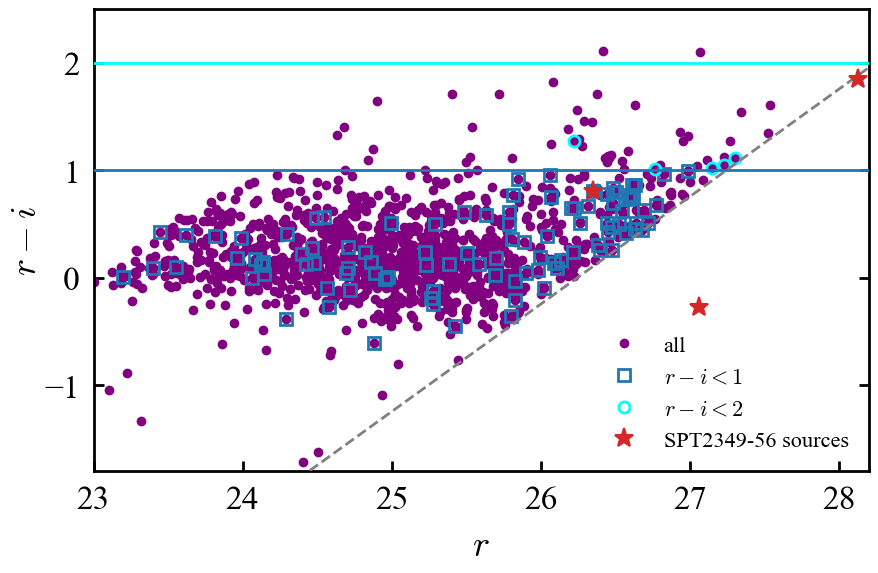}
	\includegraphics[width=0.85\textwidth]{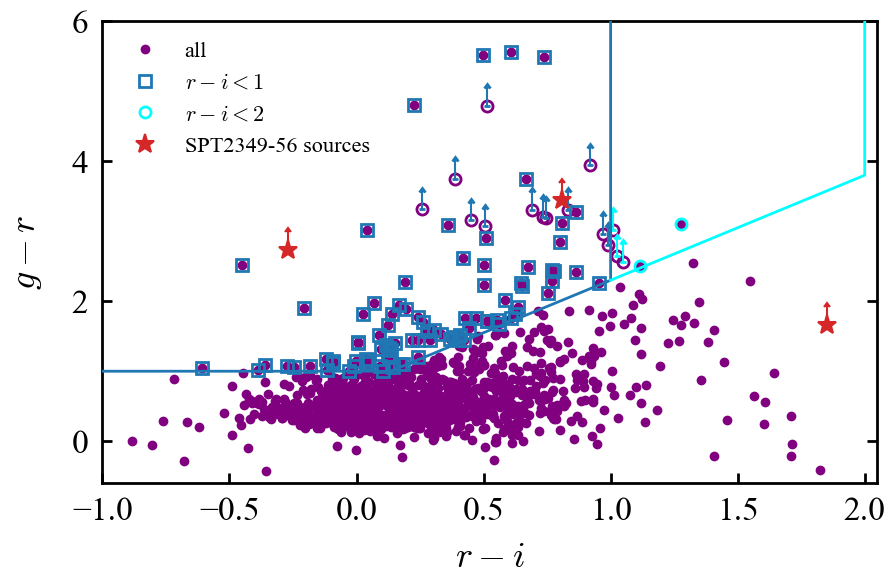}
	\caption[]{Colour-magnitude diagram {\it (top)} and colour-colour diagram {\it (bottom)} of objects selected with ${>}5\sigma$ in $i$ and detected with ${>}3\sigma$ in $r$ within the GMOS footprint (purple dots). Objects undetected in $g$-band are identified by open purple circles placed at the $g-r$ limit. The blue squares and blue lower limits highlight the LBG candidates selected by the \cite{Toshikawa18} colour criteria, and correspond to the LBGs shown in Fig.~\ref{fig:2349lbg}. The cyan circles and cyan lower limits are additional LBG candidates identified when the \cite{Toshikawa18} $r-i$ colour is relaxed to an upper limit of 2. They correspond to the cyan circles in Fig.~\ref{fig:2349lbg}. {\it Top:} The solid blue and cyan horizontal lines indicate the two $r-i$ colour limits. The dashed grey line represents the detection limit of the GMOS images ($5\sigma$ and $3\sigma$ for $i$ and $r$ band, respectively). {\it Bottom:} The solid blue line illustrates the \cite{Toshikawa18} colour-selection window, the cyan line the extension when the $r-i$ limit is relaxed to 2. {\it Both:} The three SPT2349-56 sources detected in at least one of the GMOS bands (sources $C,E$ and $M$) are included as red stars. 
	}
	\label{fig:2349CMD}
\end{figure*}

In this section we consider all
$g$-band dropouts (with detections ${>}5\sigma$ and ${>}3\sigma$ in $i$ and $r$ band, respectively) that meet the \cite{Toshikawa18} colour criteria in the approximately 6\,arcmin-diameter GMOS field surrounding SPT2349-56. 
Fig.~\ref{fig:2349CMD} shows the colours of the objects detected and illustrates the \cite{Toshikawa18} colour-selection window, highlighting the LBG candidates identified. 
The $z\simeq4$ LBGs within a 2.1\,arcmin radius from the ALMA centroid are listed in Table~\ref{table:dataLBGmag}.

\begin{table*}
\setlength{\tabcolsep}{5pt}
\caption[dum]{\small Properties of all candidate $z\simeq4$ LBGs within a 2.1\,arcmin radius of the SPT2349-56 ALMA centroid. Ordered by $i$-band magnitude. Photometry errors range from 0.02 for the brightest detections to 0.2 for the faintest.}
\label{table:dataLBGmag}\begin{threeparttable}
\centering
\begin{tabular}{lllccccc}\hline
ID &  \multicolumn{1}{c}{RA} & \multicolumn{1}{c}{Dec} & $i$ & $r$ & $g$ & $r-i$ & $g-r$ \\
\hline
{} &  &   &  [AB] &  [AB] &  [AB] & {[AB]} & {[AB]} \\
\hline
1 & 23:49:45.66 & {-}56:36:19.553 & 23.02 & 23.45 & 24.90 & 0.42 & 1.45 \\ 
2 & 23:49:47.736 & {-}56:37:39.9227 & 23.43 & 23.82 & 25.31 & 0.39 & 1.49 \\ 
3 & 23:49:57.3359 & {-}56:38:04.8559 & 23.45 & 23.55 & 24.56 & 0.10 & 1.01 \\ 
4 & 23:49:38.504 & {-}56:38:07.7406 & 23.78 & 23.96 & 25.06 & 0.18 & 1.10 \\ 
5 & 23:49:32.9717 & {-}56:37:03.6818 & 23.89 & 24.30 & 25.81 & 0.41 & 1.52 \\ 
6 & 23:49:41.3754 & {-}56:39:21.4016 & 23.91 & 24.08 & 25.18 & 0.17 & 1.10 \\ 
7 & 23:49:33.2647 & {-}56:39:50.1271 & 23.98 & 24.54 & 26.22 & 0.56 & 1.67 \\ 
8 & 23:49:36.6043 & {-}56:39:56.5279 & 24.01 & 24.13 & 25.50 & 0.12 & 1.37 \\ 
9 & 23:49:36.7826 & {-}56:39:42.2089 & 24.11 & 24.15 & 25.32 & 0.04 & 1.17 \\ 
10 & 23:49:50.3222 & {-}56:38:46.3643 & 24.18 & 24.40 & 25.84 & 0.22 & 1.44 \\ 
11 & 23:49:48.1164 & {-}56:36:36.9346 & 24.42 & 24.71 & 26.16 & 0.29 & 1.45 \\ 
12 & 23:49:44.8061 & {-}56:36:49.2048 & 24.48 & 25.00 & - & 0.51 & - \\ 
13 & 23:49:45.9239 & {-}56:37:09.4512 & 24.79 & 25.29 & - & 0.50 & - \\ 
14 & 23:49:42.4754 & {-}56:36:42.8929 & 24.84 & 24.89 & 26.01 & 0.04 & 1.12 \\ 
15 & 23:49:39.1115 & {-}56:39:33.6838 & 24.87 & 25.48 & - & 0.61 & - \\ 
16 & 23:49:52.0815 & {-}56:38:29.886 & 24.93 & 25.85 & - & 0.92 & - \\ 
17 & 23:49:53.935 & {-}56:37:22.4742 & 24.95 & 26.22 & 29.33 & 1.28 & 3.11 \\ 
18 & 23:49:32.9176 & {-}56:38:37.2422 & 24.97 & 24.96 & 26.03 & -0.012 & 1.07 \\ 
19 & 23:49:48.3729 & {-}56:37:24.8416 & 24.97 & 24.98 & 26.39 & 0.007 & 1.42 \\ 
20 & 23:49:32.1277 & {-}56:37:52.5108 & 24.98 & 25.23 & 27.00 & 0.24 & 1.78 \\ 
21 & 23:49:46.3815 & {-}56:37:33.352 & 25.05 & 25.64 & 27.65 & 0.58 & 2.02 \\ 
22 & 23:49:47.4104 & {-}56:37:48.3888 & 25.11 & 26.06 & 28.32 & 0.95 & 2.26 \\ 
23 & 23:49:42.884 & {-}56:37:41.5758 & 25.18 & 25.79 & 27.55 & 0.61 & 1.76 \\ 
24 & 23:49:45.4183 & {-}56:37:50.6935 & 25.26 & 25.39 & 26.61 & 0.13 & 1.22 \\ 
25 & 23:49:41.2714 & {-}56:39:24.5941 & 25.28 & 25.51 & - & 0.23 & - \\ 
26 & 23:49:50.0852 & {-}56:38:37.3747 & 25.45 & 25.81 & 28.90 & 0.36 & 3.09 \\ 
27 & 23:49:52.6003 & {-}56:38:16.6182 & 25.46 & 25.59 & 26.92 & 0.13 & 1.33 \\ 
28 & 23:49:48.6547 & {-}56:37:05.4282 & 25.49 & 24.88 & 25.92 & -0.607 & 1.04 \\ 
29 & 23:49:43.3442 & {-}56:38:20.9623 & 25.54 & 26.35 & 29.46 & 0.81 & 3.11 \\ 
30 & 23:49:51.6146 & {-}56:39:22.7974 & 25.65 & 26.48 & - & 0.83 & - \\ 
31 & 23:49:37.0779 & {-}56:39:35.7322 & 25.65 & 26.32 & - & 0.67 & - \\ 
32 & 23:49:53.3103 & {-}56:38:23.5745 & 25.67 & 26.45 & 28.88 & 0.78 & 2.43 \\ 
33 & 23:49:52.7764 & {-}56:39:42.9613 & 25.70 & 26.50 & 29.35 & 0.80 & 2.85 \\ 
34 & 23:49:52.535 & {-}56:37:20.5525 & 25.75 & 26.62 & - & 0.86 & - \\ 
35 & 23:49:42.6368 & {-}56:38:29.0018 & 25.76 & 26.77 & - & 1.01 & - \\ 
36 & 23:49:38.5943 & {-}56:38:44.4538 & 25.77 & 26.63 & 29.06 & 0.86 & 2.42 \\ 
37 & 23:49:38.4095 & {-}56:36:35.3473 & 25.79 & 26.48 & - & 0.69 & - \\ 
38 & 23:49:41.5607 & {-}56:37:23.0682 & 25.83 & 26.60 & 28.90 & 0.77 & 2.30 \\ 
39 & 23:49:32.0454 & {-}56:37:46.6313 & 25.86 & 25.83 & 26.83 & -0.030 & 1.00 \\ 
40 & 23:49:43.2724 & {-}56:38:29.7938 & 25.86 & 26.83 & - & 0.97 & - \\ 
41 & 23:49:39.8886 & {-}56:37:33.7577 & 25.86 & 26.61 & - & 0.75 & - \\ 
42 & 23:49:51.4217 & {-}56:38:53.7364 & 25.88 & 26.62 & - & 0.74 & - \\ 
43 & 23:49:45.9803 & {-}56:36:31.3668 & 25.91 & 25.98 & 27.96 & 0.07 & 1.98 \\ 
44 & 23:49:38.8815 & {-}56:39:14.4601 & 25.94 & 26.45 & 28.17 & 0.51 & 1.72 \\ 
45 & 23:49:53.3111 & {-}56:37:56.4834 & 25.95 & 26.09 & 27.90 & 0.14 & 1.81 \\ 
46 & 23:49:57.1831 & {-}56:38:14.7836 & 25.98 & 26.22 & 27.66 & 0.24 & 1.44 \\ 
47 & 23:49:39.3771 & {-}56:39:12.8362 & 26.01 & 26.48 & 28.23 & 0.47 & 1.75 \\ 
48 & 23:49:33.3948 & {-}56:37:27.323 & 26.02 & 26.11 & 27.63 & 0.09 & 1.52 \\ 
49 & 23:49:41.9175 & {-}56:36:26.9575 & 26.02 & 26.45 & 28.21 & 0.43 & 1.76 \\ 
50 & 23:49:36.8624 & {-}56:38:00.424 & 26.03 & 26.53 & 28.76 & 0.50 & 2.23 \\ 
51 & 23:49:34.931 & {-}56:38:55.1054 & 26.08 & 26.38 & 27.98 & 0.31 & 1.59 \\ 
52 & 23:49:30.1492 & {-}56:38:18.3113 & 26.10 & 26.77 & 29.25 & 0.67 & 2.48 \\ 
53 & 23:49:46.7053 & {-}56:36:35.3912 & 26.12 & 27.15 & - & 1.03 & - \\ 
54 & 23:49:31.3236 & {-}56:37:06.0532 & 26.13 & 26.64 & 29.54 & 0.51 & 2.90 \\ 
55 & 23:49:31.7981 & {-}56:38:39.521 & 26.14 & 26.40 & 28.10 & 0.26 & 1.71 \\ 
56 & 23:49:49.3724 & {-}56:37:00.4062 & 26.16 & 25.80 & 26.89 & -0.359 & 1.09 \\ 
57 & 23:49:46.8108 & {-}56:36:35.4287 & 26.22 & 26.72 & - & 0.51 & - \\ 
\hline\end{tabular} 
\end{threeparttable}
\end{table*}

\bsp	
\label{lastpage}
\end{document}